\documentclass[twocolumn,superscriptaddress,preprintnumbers,amsmath,amssymb,longbibliography,floatfix]{revtex4-1}
\usepackage{amsmath}
\usepackage{amssymb}
\usepackage{dcolumn}
\usepackage{graphicx}
\usepackage{subfigure}
\usepackage[american]{circuitikz}
\usepackage{tikz}
\usepackage{bm}
\usepackage{epstopdf}
\usepackage{threeparttable}
\usepackage{float}
\usepackage{color}
\usepackage{mathtools} 
\usepackage{mathtools}

\begin{document}
\title{Unifying the Concepts of Scattering  and  Structure Factor in  Ordered and Disordered Samples}
\author{Dingning Li} 
\affiliation{Division of Natural and Applied Sciences, Duke Kunshan University, Kunshan, Jiangsu, 215300, China}
\author{Kai Zhang} 
\email{kai.zhang@dukekunshan.edu.cn}
\affiliation{Division of Natural and Applied Sciences, Duke Kunshan University, Kunshan, Jiangsu, 215300, China}

\date{\today}

\begin{abstract}
Scattering methods are widely used in many research areas to analyze and resolve material structures. Given the importance, a large number of full textbooks are  devoted to this topic. However, technical details in experiments and disconnection between   explanations from different perspectives  often confuse and frustrate beginner students and researchers.  To  create an effective learning path, we  review the core concepts about scattering and structure factor in this article in a  self-contained way.  Classical examples of scattering photography and spectroscopy are calculated. Sample CPU and GPU codes are provided to facilitate the understanding and application of these methods.
\end{abstract}
 
\maketitle

\section{Introduction}
 
 Scattering methods,  using a  source of  photons, electrons, X-rays, neutrons, etc.,  are powerful tools to examine microscopic structural~\cite{powles1973} and dynamical~\cite{goldburg1999} properties of matter, which have been successfully  applied to study  subatomic particles~\cite{xiong2019}, crystals~\cite{azaroff1968},  liquids~\cite{head2002}, glasses~\cite{sette1998}, surfactants~\cite{hayter1983},  biomolecules~\cite{kendrew1961, ashkar2018} and polymers~\cite{roe2000}.  The rule of thumb here  is that the wavelength $\lambda$ of the radiation  should be comparable to the length scale of the structure to be observed.   To detect  ordering over a range much longer than $\lambda$, methods like  small-angle scattering (SAS)  are needed~\cite{chu2001}. Another important consideration is about the contrast between scattering signals from different elements due to underlying physical mechanisms. Therefore, neutron scattering is often  preferred for soft-matter systems, despite its lower accessibility than for X-rays.  In addition,  techniques like resonant soft-X-ray scattering  can be   used  to provide  enhanced   resolution~\cite{fink2013,liu2016}. Compared with real-space microscopy techniques, reciprocal-space probes like scattering methods are good at picking up periodic patterns and revealing three-dimensional structures as a whole by penetrating deeply into the sample~\cite{mukherjee2017}.

Given the richness of material structures,  a variety of experimental methods have been developed during the last century, with the scattering being  hard (high-energy) or soft (low-energy),   monochromatic or  polychromatic,    elastic  or inelastic. Despite the diversity of   experimental setups,  they can largely be grouped into two categories based on how signals are collected and interpreted. The first category is {\em photography} of   ordered samples,  which  are recorded  as spotted  scattering signals  on a two-dimensional  (2D)   film~\cite{mcintyre2015}. The second category is {\em intensity scanning}   of scattering  signals   from  disordered or partially ordered samples,   whose  one-dimensional (1D) profile is plotted against one variable (a scalar)  that characterizes the existence of periodicities  in the system~\cite{hura2009}. In both types, the quantitative measurement of the  signal  is  scattering intensity $I({\bf q})$,  or its normalized version, structure factor $S({\bf q})$, which is often expressed as a function of  scattering vector ${\bf q}$.
 The central tasks of structural analysis  with scattering methods  are then
\begin{itemize}
\item the forward problem $\rho({\bf r})\rightarrow I({\bf q}) $:  given the electron  density distribution    $\rho({\bf r})$ or particle positions $({\bf r}_1, {\bf r}_2, \cdots, {\bf r}_N)$,  to predict the scattering pattern $I({\bf q})$; and
\item the inverse problem  $  I({\bf q}) \rightarrow  \rho({\bf r}) $: given the scattering pattern $I({\bf q})$, to resolve the electron  density distribution    $\rho({\bf r})$ or particle positions $({\bf r}_1, {\bf r}_2, \cdots, {\bf r}_N)$. 
\end{itemize}
In this article, we only focus on the forward problem, which could still shed light on some basic structural information. Sometimes, the forward method may  also be used to solve $\rho({\bf r})$ iteratively, through a trial-and-error process. That is, one keeps modifying a proposed structure   $\rho({\bf r})$ until the theoretically computed  $I({\bf q})$ matches the experimentally observed one.
The full solution to the  inverse problem is, however,  challenged by the  notorious  ``phase problem''~\cite{hauptman1991}.

Concepts about scattering and structure factor are often discussed across different disciplines including condensed-matter physics, materials science, polymer physics, structure biology, etc.  The same idea  can take different forms in different areas, causing confusions and  misconceptions.    Graduate or advanced undergraduate students   in need of applying these concepts to their research problems  can be frustrated by  the convoluted  experimental details covered in traditional textbooks. 
It is thus the purpose of this article to unify the concepts about scattering and structure factor,  giving junior researchers   an effective pathway to quickly grasp the key ideas  in this field without taking a whole course or reading an entire textbook.

 To fulfill this task, we first elaborate the fundamentals about scattering (Section~\ref{sec:scatter}), crystallography (Section~\ref{sec:crystal}) and liquid-state theory (Section~\ref{sec:liquid}) based on the Fourier transform and reciprocal lattice.  Using concrete examples, we then discuss the photography of ordered samples in Section~\ref{sec:exp}-\ref{sec:photo} as well as intensity scanning of isotropic samples in Section~\ref{sec:q}-\ref{sec:spec}.  Relevant  CPU and GPU source codes  are provided online at \url{https://github.com/statisticalmechanics/scatter}. Finally, a brief introduction to the 2D structure factor is given in Section~\ref{sec:2Dsq}, before the conclusion in Section~\ref{sec:con}.

\section{Scattering}

\label{sec:scatter}

\subsection{Scattering Vector }

\begin{figure}
\includegraphics[width=0.45\textwidth]{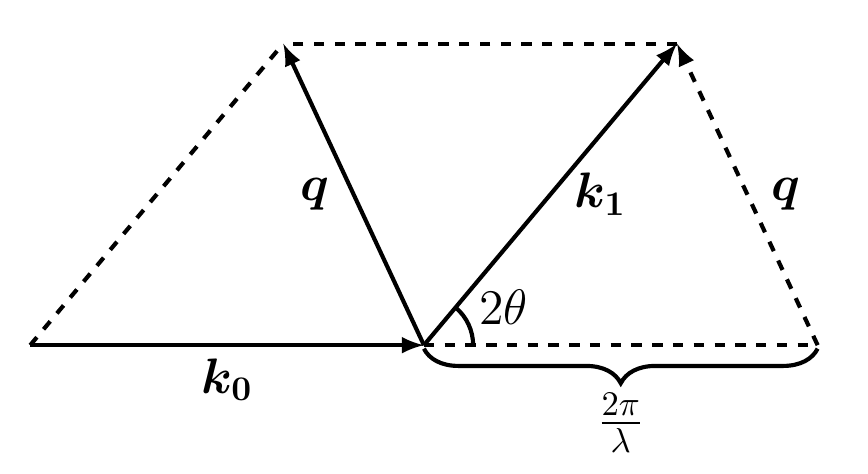}
\caption{Scattering vector ${\bf q}$ defined as the difference between the diffracted wavevector ${\bf k}_1$ and the incident wavevector ${\bf k}_0$,  both with magnitude  $2\pi / \lambda$ during elastic scattering.}
\label{fig:k0k1}
\end{figure}

In a scattering experiment, the incident beam  of wavevector ${\bf k}_0$, after hitting the sample,  is deflected from its straight path  by a {\em scattering angle} $2\theta$ and becomes the diffracted beam of wavevector ${\bf k}_1$ (Fig.~\ref{fig:k0k1}).  In  case of elastic~\footnote{It should be noted that the diffraction experiment generally detects both elastic and inelastic scattering, where the latter results from dynamic processes in the sample. To measure just elastic scattering, an energy analyser should be placed between the sample and detector.} and monochromatic scattering  (of a fixed wavelength $\lambda$), $ |{\bf k}_0|= |{\bf k}_1| = \frac{2\pi}{\lambda}$. The change of wavevector,  called the
{\em scattering vector},  is 
\begin{align}
{\bf q} = {\bf k}_1 - {\bf k}_0
\end{align}
with a magnitude
 \begin{align}
 \label{eq:qtheta}
q = 2 |{\bf k}_0| \sin \theta = \frac{4 \pi}{\lambda} \sin \theta.
\end{align}
Let  ${\bf s}_0 =  {\bf k}_0  / |{\bf k}_0 | = {\bf k}_0 \lambda / (2\pi ) $ and ${\bf s}_1 =  {\bf k}_1  / |{\bf k}_1 | ={\bf k}_1 \lambda / ( 2\pi )$ be the unit vectors of the incident and diffracted beam, then the scattering vector can also be written as
 \begin{align}
 \label{eq:qs}
 {\bf q}  = \frac{2\pi}{\lambda} ({\bf s}_1 - {\bf s}_0).
\end{align}

\subsection{Scattering Intensity}

When a detection  screen is placed behind the sample in the path of ${\bf k}_1$,  the diffracted beam  may be detected. The strength of such signals is quantified by the {\em scattering intensity}  $I({\bf q})$ of the ray, which changes with ${\bf k}_1$, or equivalently,  with ${\bf q}$. 
The scattering pattern, or the distribution of $I({\bf q})$ on the   screen, is determined by the structural features of the sample, for instance,   the electron  density distribution $\rho({\bf r})$ in the case  of X-ray scattering by atoms.

Both the incident and the diffracted rays can be viewed as plane waves of the form $ \psi_{{\bf  k }}({\bf  r })  = \langle {\bf  r }| {\bf k} \rangle \propto  e^{ i {\bf k} \cdot {\bf r}}  $. According to  Fermi's golden rule, the  scattering  intensity $I({\bf q})$  is proportional to the square of the transition probability amplitude from state  $ \psi_{{\bf  k}_0}({\bf  r }) $ to  state $ \psi_{{\bf  k}_1}({\bf  r }) $,  after interacting with the overall scattering potential  $\rho({\bf r}) $.  That is,
\begin{align*}
\begin{array}{ll}
I({\bf q}) &\propto \left |    \langle  {\bf k}_0 |\rho({\bf r})  |  {\bf k}_1 \rangle   \right |^2  =  \left |   \int   d{\bf r}    \psi^*_{{\bf  k}_0}({\bf  r })    \rho({\bf r})     \psi_{{\bf  k}_1}({\bf  r })  \right |^2 \\
& \propto  \left |     \int   d{\bf r}    e^{ - i {\bf k}_0 \cdot {\bf r}}   \rho({\bf r})     e^{ i {\bf k}_1 \cdot {\bf r}}  \right |^2 \\
&=  \left |  \int   d{\bf r}     \rho({\bf r})  e^{ i {\bf q} \cdot {\bf r}}   \right |^2 
\end{array}
\end{align*}
Neglecting the coefficient of proportionality, one can write
\begin{align}
I({\bf q})  =  \hat{\rho}_{\bf q} \hat{\rho}_{\bf -q},
\end{align}
where
\begin{align}
\hat{\rho}_{\bf q} = \int  d{\bf r}  \rho({\bf r})  e^{ i {\bf q} \cdot {\bf r}}  
\end{align}
is the Fourier transform of the density distribution and  $\hat{\rho}_{\bf -q}$ is its complex conjugate (Appendix~\ref{sec:ft}). 

Unless $ \rho({\bf r}) $ has  a symmetry center,  $\hat{\rho}_{\bf q}$ is generally a complex number, i.e.  $\hat{\rho}_{\bf q} = |\hat{\rho}_{\bf q}| e^{i \phi_{{\bf q}}}$. If $\hat{\rho}_{\bf q} $ is known exactly, $ \rho({\bf r}) $  can  in principle be  reconstructed through the inverse Fourier transform Eq.~(\ref{eq:fourinv})~\cite{argos1977}. However, in an experiment, only the scattering intensity  $ I({\bf q})    = |\hat{\rho}_{\bf q}|^2     e^{i \phi_{{\bf q}}}  e^{- i \phi_{{\bf q}}} =  |\hat{\rho}_{\bf q}|^2  $  is directly measurable. This allows us to compute the magnitude of  $\hat{\rho}_{\bf q}$ by  $|\hat{\rho}_{\bf q}|  = \sqrt{ I({\bf q})  }$.  Unfortunately, information about the phase angle $\phi_{{\bf q}}$  is lost during this process, which gives rise to the 
``phase problem'' in crystallography.   Special techniques~\cite{hauptman1991,harrison1993,taylor2003} have been developed to determine   $\phi_{{\bf q}}$,  which are beyond the scope of this article.

In a system of $N$ atoms or particles at positions $({\bf r}_1, {\bf r}_2, \cdots, {\bf r}_N)$ inside a region of volume $V$, 
the density distribution  consists of the contributions from each particle $i$  with a scattering potential $f_i( {\bf r} - {\bf r}_i)$ ($i=1,2,\cdots,N$), i.e.
\begin{align}
\rho({\bf r}) = \sum_{i=1}^N  f_i( {\bf r} - {\bf r}_i) =   \sum_{i=1}^N  f_i({\bf R}_i),~~({\bf R}_i\equiv {\bf r} - {\bf r}_i).
\end{align}
 In this case
 \begin{align}
 \label{eq:rhoq}
\begin{array}{ll}
\hat{\rho}_{\bf q}& =      \int\limits_V   d{\bf r}    \sum\limits_{i=1}^N  f_i( {\bf r} - {\bf r}_i)  e^{ i {\bf q} \cdot {\bf r}}    \\
& =     \sum\limits_{i=1}^N   \int\limits_V   d{\bf R}_i     f_i( {\bf R}_i)  e^{ i {\bf q} \cdot {\bf R}_i}  e^{ i {\bf q} \cdot {\bf r}_i}     \\
& =     \sum\limits_{i=1}^N      \hat{f}_i( {\bf q})    e^{ i {\bf q} \cdot {\bf r}_i}    
\end{array}
\end{align}
where  
 \begin{align}
\hat{f}_i( {\bf q})  =     \int\limits_V   d{\bf r}     f_i( {\bf r})  e^{ i {\bf q} \cdot {\bf r}}  
\end{align}
is the {\em atomic form factor}, or scattering factor, of particle $i$.

 If the scattering potential of each particle $f_i( {\bf r} - {\bf r}_i)$ is symmetric about ${\bf r}_i$, which should be true for atoms and most particles,   $\hat{f}_i( {\bf q})$ is  \underline{real and even}, i.e. its complex conjugate, $\hat{f}^*_i( {\bf q}) = \hat{f}_i( {\bf -q})  =   \hat{f}_i( {\bf q}) $ (Appendix~\ref{sec:ft}) 
Under this circumstance, the scattering intensity
\begin{align}
 \label{eq:Iq}
\begin{array}{ll}
I({\bf q})  & =       \sum\limits_{i=1}^N    \hat{f}_i( {\bf q})   e^{ i {\bf q} \cdot {\bf r}_i}     \sum\limits_{j=1}^N    \hat{f}_j( -{\bf q})   e^{- i {\bf q} \cdot {\bf r}_j}    \\
& =       \sum\limits_{i=1}^N    \hat{f}_i( {\bf q})   e^{ i {\bf q} \cdot {\bf r}_i}     \sum\limits_{j=1}^N    \hat{f}_j( {\bf q})   e^{- i {\bf q} \cdot {\bf r}_j}    \\
& = \left |      \sum\limits_{i=1}^N \hat{f}_i( {\bf q})   \cos({\bf q} \cdot {\bf r}_i ) \right|^2  +  \left |      \sum\limits_{i=1}^N  \hat{f}_i( {\bf q})  \sin({\bf q} \cdot {\bf r}_i ) \right|^2 
\end{array}
\end{align}
or equivalently, 
\begin{align}
 \label{eq:Iq_ij}
\begin{array}{ll}
I({\bf q})  &  =   \sum\limits_{i=1}^N    \sum\limits_{j=1}^N  \hat{f}_i( {\bf q})  \hat{f}_j( {\bf q})   e^{ i {\bf q} \cdot ( {\bf r}_i - {\bf r}_j  ) }  \\
& =  \sum\limits_{i=1}^N    \sum\limits_{j=1}^N   \hat{f}_i( {\bf q})  \hat{f}_j( {\bf q})  e^{ i {\bf q} \cdot  {\bf r}_{ij} }   \\
&  =   \sum\limits_{i=1}^N    \sum\limits_{j=1}^N  \hat{f}_i( {\bf q})  \hat{f}_j( {\bf q})  \cos ( {\bf q} \cdot  {\bf r}_{ij} )  \\
&  =   \sum\limits_{i=1}^N  \hat{f}_i^2( {\bf q})  + \sum\limits_{i=1}^N    \sum\limits_{j \neq i }^N  \hat{f}_i( {\bf q})  \hat{f}_j( {\bf q})  \cos ( {\bf q} \cdot  {\bf r}_{ij} ).
\end{array}
\end{align}
Eq.~(\ref{eq:Iq}) and Eq.~(\ref{eq:Iq_ij})  are mathematically equivalent because  $\cos ( {\bf q} \cdot  {\bf r}_{ij} )  = \cos ( {\bf q} \cdot    {\bf r}_i -  {\bf q} \cdot   {\bf r}_j    )   =   \cos ( {\bf q} \cdot    {\bf r}_i )  \cos(  {\bf q} \cdot   {\bf r}_j    )    +  \sin ( {\bf q} \cdot    {\bf r}_i )  \sin(  {\bf q} \cdot   {\bf r}_j    )  $. However, in numerical computation of  $I({\bf q})$  at a given ${\bf q}$,  Eq.~(\ref{eq:Iq}) has a lower cost with a computational complexity   ${\mathcal O}(N)$, while Eq.~(\ref{eq:Iq_ij})  is of  complexity  ${\mathcal O}(N^2)$. Nevertheless, when there is an appropriate symmetry in the system, the expression $ {\bf r}_{ij} $ in   Eq.~(\ref{eq:Iq_ij})  allows it to be further simplified and thus to become  more computationally efficient, as will be discussed in later sections.

\subsection{Atomic Form Factor}

\begin{figure}
\centering
\includegraphics[width=0.45\textwidth]{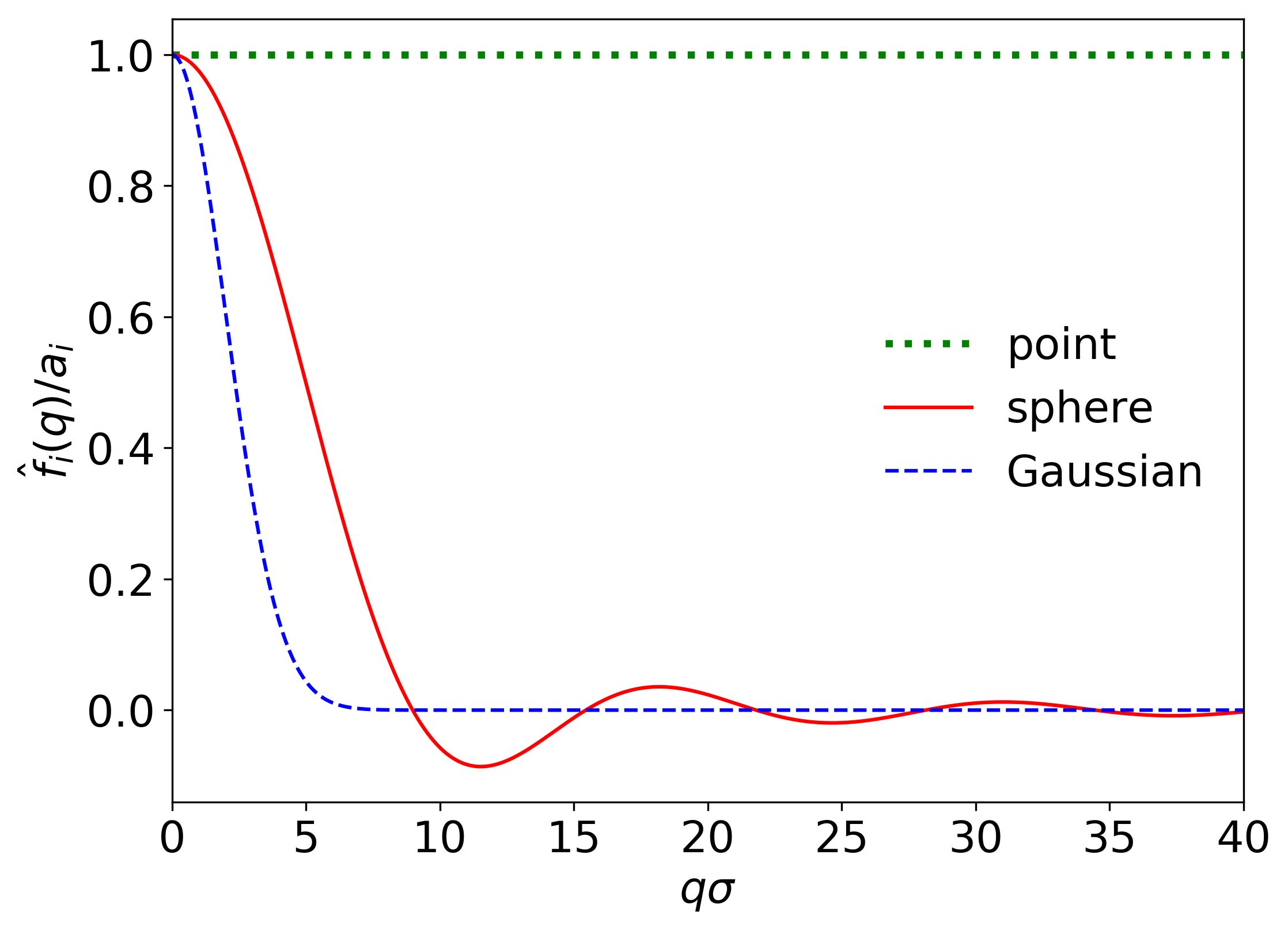}
\caption{Atomic form factor $\hat{f}_i(q)$ of a sizeless point  (green dotted, Eq.~\ref{eq:fpoint}),  a uniform spherical (red solid, Eq.~\ref{eq:sphere}) and a Gaussian scattering center (blue dashed, Eq.~\ref{eq:gaussian}) as a function of $q$. }
\label{fig:fq}
\end{figure}

 For realistic scattering potentials, the atomic form factor $\hat{f}_i( {\bf q}) $  changes with the direction and magnitude of the  scattering vector ${\bf q}$, and thus often drops as the scattering angle $\theta$ increases (Fig.~\ref{fig:fq}). 
If, however, the scattering potential is  \underline{spherically symmetric}, i.e. $f_i({\bf r}) = f_i(r)$, we can write
 \begin{align}
  \begin{aligned}
\hat{f}_i( {\bf q})  &=  \hat{f}_i( q)   = 2\pi   \int   dr  r^2    f_i( r )   \int_0^{\pi} d\theta  \sin \theta    e^{ i q r \cos \theta}   \\
& = 2\pi   \int   dr  r^2    f_i( r )  \frac{2\sin (q r)}{qr} \\
& = 4\pi   \int   dr  r^2    f_i( r )  \frac{\sin (q r)}{qr}~~~(q\ne 0).\\
\end{aligned}
\end{align}
It is useful to consider the three simple spherically symmetric scattering potentials listed below (Fig.~\ref{fig:fq}).
\begin{itemize}
\item  $f_i( {\bf r} - {\bf r}_i) = a_i  \delta ( {\bf r} -  {\bf r}_i )  $,  the scattering by each atom is idealized as from a sizeless point at the atomic center. This model can be mapped onto the physical scenario of nuclear scattering or the abstract scenario of point-mass scattering. The scattering strength $a_i$  of atom $i$   generally has  different values  for different elements, which has also been called  the atomic scattering factor, because  here  
\begin{align}
\label{eq:fpoint}
\hat{f}_i( {\bf q}) = a_i.
\end{align}
The electron density distribution  is then $\rho({\bf r}) =  \sum\limits_{i=1}^N a_i  \delta ( {\bf r} -  {\bf r}_i ) $, which,  in the case of $a_i = 1$,   becomes the particle density distribution $\rho({\bf r}) =  \sum\limits_{i=1}^N  \delta ( {\bf r} -  {\bf r}_i ) $.

\item  $ f_i( {\bf r} - {\bf r}_i)$ is homogeneous and bounded within a sphere of radius $\sigma/2$~\footnote{Throughout the paper, we use $\sigma$ as the unit of length and $1/\sigma$ as the unit of wavevector. We  choose $\sigma$ to be particle diameter, which can be mapped onto the length scale of  ${\rm \AA}$  for atomic systems, nm for nano-systems, and $\mu$m for colloidal systems.  }, 
 \begin{align}
 f_i  ( {\bf r} - {\bf r}_i) = \left \{
 \begin{array}{ll}
&  a_i / (\pi \sigma^3 /6),  ~~ | {\bf r} - {\bf r}_i| \le \sigma/2 \\
& 0, ~~~{\rm otherwise}
\end{array}
\right .
\end{align}
and
 \begin{align}
 \label{eq:sphere}
 \begin{aligned}
  \hat{f}_i(q) & = \frac{4\pi a_i /(\pi \sigma^3 /6)}{q^3} [ \sin(q \sigma/2)  - q \sigma/2  \cos(q \sigma /2)]  \\
 &  = \frac{3  a_i  }{(q\sigma/2)^3} [ \sin(q \sigma/2)  - q \sigma/2  \cos(q \sigma /2)] .
 \end{aligned}
 \end{align}

\item   $ f_i( {\bf r} - {\bf r}_i)$ is Gaussian-like with standard deviation $\sigma/2$,
 \begin{align}
 \begin{aligned}
  f_i( {\bf r} - {\bf r}_i) &= a _i\left(\frac{1}{\sqrt{2\pi(\sigma/2)^2}}\right)^3 e^{-\frac{| {\bf r} - {\bf r}_i|^2}{2(\sigma/2)^2}} \\
  &=  \frac{a_i}{\sigma^3 (\pi/2)^{3/2}} e^{- \frac{2 R_i^2}{\sigma^2}} ~~ (R_i =  |{\bf r} - {\bf r}_i|)
 \end{aligned}
  \end{align}
 and
 \begin{align}
 \label{eq:gaussian}
 \hat{f}_i( q) =  a_i e^{- \frac{\sigma^2 q^2}{8}}.
  \end{align}
\end{itemize}
In all numerical results shown below, we will assume $\hat{f}_i( {\bf q}) = 1$, i.e. point scattering,  for all particles.

\section{Crystallography}

\label{sec:crystal}
We now  review concepts and theories about scattering methods used for crystal samples. The earlier theory of von Laue~\cite{mcquarrie1997} that considers diffraction of parallel beams by arrays of atoms is skipped here. Instead, we apply the more intuitive Bragg's law that envisages crystallographic planes as reflective mirrors to understand the principle, although there is no such reflection in the physical sense.

\subsection{Bragg's Law}

\begin{figure}
\centering
\includegraphics[width=0.45\textwidth]{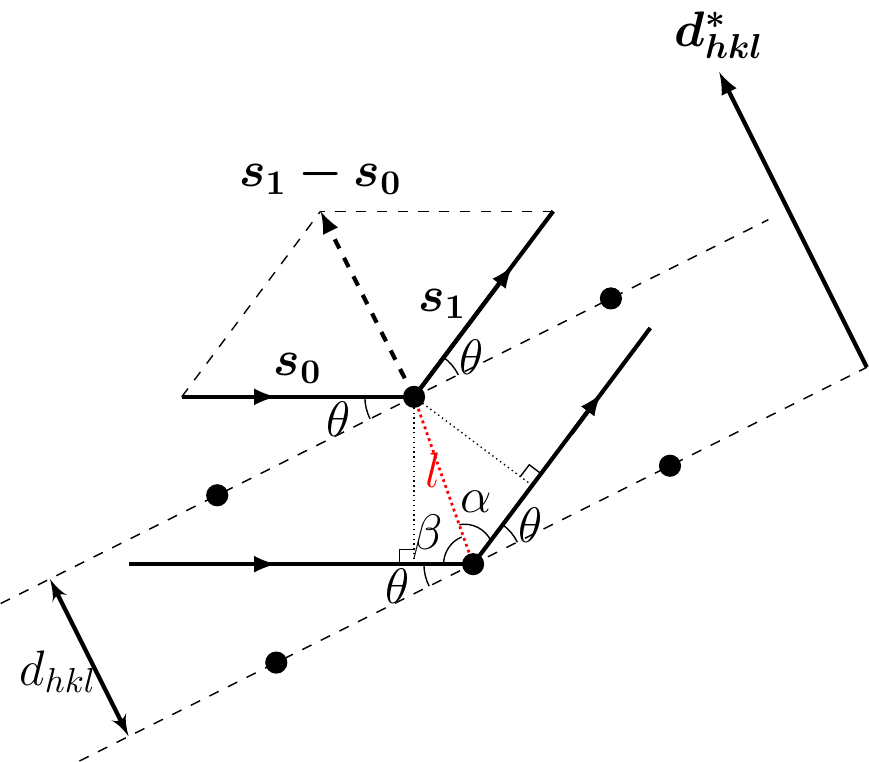}
\caption{The scattering paths of two  rays diffracted by two layers of ordered particles (black dots) with interplanar distance $d_{hkl}$ and scattering angle $2\theta$.  ${\bf s}_0$ and ${\bf s}_1$ are unit vectors of the incident and diffracted rays. When  Bragg's law is satisfied, the scattering vector is parallel to the  normal vector  ${\bf d}_{hkl} ^*$ of the lattice planes.}
\label{fig:bragg}
\end{figure}

For an incident ray of wavelength $\lambda$  to generate a strong constructive scattering signal by a family of crystallographic planes $(hkl)$ of interplanar spacing $d_{hkl}$ (Appendix~\ref{sec:lat}),  the scattering angle $2\theta$ needs to obey  Bragg's law~\cite{bragg1968} (Fig.~\ref{fig:bragg})
\begin{equation}\label{eq:bragg}
n\lambda=2d_{hkl}\sin\theta,~~n=1,2,3,\cdots.
\end{equation}
This is because the path difference of the two scattering rays ``reflected''  by two neighboring lattice planes is 
 \begin{align*}
 \begin{aligned}
l (\cos \alpha + \cos \beta) & = l [ \cos \alpha + \cos (\pi - 2\theta - \alpha) ] \\
& = l [ \cos \alpha -  \cos ( 2\theta + \alpha) ] \\
& = l  \sin (\theta + \alpha) \sin \theta \\
& = d_{hkl} \sin \theta.
 \end{aligned}
  \end{align*}
  
  The rescaled scattering vector ${\bf s}_1 - {\bf s}_0 = \frac{\lambda}{2\pi} {\bf q}$ (of  length  $2\sin\theta$)  is parallel to the normal vector, or reciprocal vector ${\bf  d}^*_{hkl}$ (of length $1 / d_{hkl}$), of the lattice planes $(hkl)$. Thus, it is sometimes  convenient to express Bragg's law in a vector form, for instance, for the  primary $n=1$ scattering, as
  \begin{align}\label{eq:bragg_vec}
  \frac{{\bf s}_1 - {\bf s}_0}{\lambda}  =  {\bf d}^*_{hkl} .
\end{align}
Using Eq.~(\ref{eq:qs}), the necessary condition to receive a strong signal for scattering vector ${\bf q}$ in crystals is thus
  \begin{align}\label{eq:bragg_q}
  {\bf q} =  2\pi {\bf d}^*_{hkl} .
\end{align}

\subsection{The Ewald Construction}

Bragg's law  needs to be satisfied to have a strong  scattering signal    in the direction of ${\bf s}_1$. However, this does not mean that, given an arbitrary experimental setup, Bragg's law  is guaranteed to be satisfied somewhere.  In particular, if a monochromatic incident beam (fixed $\lambda$) is directed onto a single crystal at an arbitrarily  fixed position (fixed $\theta$'s and $d_{hkl}$'s), it is possible that none of the lattice planes will be able to produce a strong scattering signal. If this happens, either $\lambda$ (polychromatic) or $\theta$ (rotate the sample or use polycrystals) has to been tuned to  satisfy  Eq.~(\ref{eq:bragg}-\ref{eq:bragg_q}).

\begin{figure}
\centering
\includegraphics[width=0.45\textwidth]{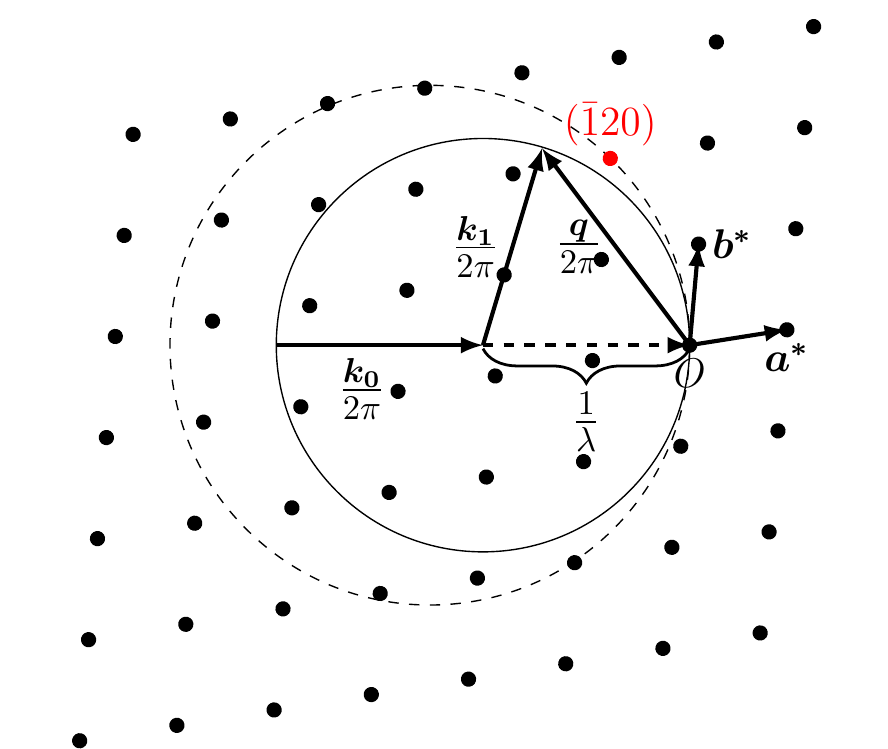}
\caption{The Ewald construction: Ewald's sphere of radius $1/\lambda$ (solid circle) depicts all possible scattering wavevectors ${\bf q}$ under the current setup.  Lattice planes with Miller indices $(hkl)$ are represented by points on the reciprocal lattice (black dots).   For wavelength $\lambda$,  no  reciprocal lattice  point is on  Ewald's sphere implying that no scattering signal will be generated at any scattering angle. If the wavelength is appropriately tuned,  some reciprocal lattice points can fall on the new Ewald's sphere (dashed circle) to satisfy   Bragg's law, for instance, $(\bar{1}20)$. }
\label{fig:ewald}
\end{figure}

An alternative view to check the satisfaction of Bragg's law is to use {\em Ewald's sphere} in the reciprocal space~\cite{hammond2001,barbour2018}. Here, each point at vector ${\bf d}_{hkl}^*$ represents a family of parallel planes $(hkl)$ in the direct space. When the orientation of  the crystal sample is fixed, the relative position of the incident beam and reciprocal lattice points are also fixed. One can align the end point of the incident wavevector ${\bf k}_0$ (in practice   ${\bf k}_0/2\pi$) with the origin $O$ of the reciprocal lattice, then draw a sphere of radius $1/\lambda$. The center of the sphere is found by moving from point $O$ by a vector  displacement $-{\bf k}_0/2\pi$ (Fig.~\ref{fig:ewald}). It can be seen that the end point of the scattering vector ${\bf q}$, normalized by $2\pi$, falls on the surface of this Ewald's sphere. According to the vector form of Bragg's law Eq.~(\ref{eq:bragg_q}), a   scattering  from  certain lattice planes $(hkl)$  is possible only when the corresponding reciprocal vector point ${\bf d}_{hkl}^*$  falls on the surface of   Ewald's sphere. If wavelength  and crystal orientation  are not appropriately chosen,   this condition may not  be met at all and no scattering signal is generated by the sample.

\subsection{Crystal Structure Factor  $F_{hkl}$}

\label{sec:Fhkl}
Bragg's law  is actually the  necessary (but not sufficient) condition to have a strong  scattering signal.
Even if Bragg's law  is obeyed by lattice planes $(hkl)$, it is still possible that the scattering signal cancels due to   special lattice symmetries.  In fact, when  Bragg's law is presented as in Fig.~\ref{fig:bragg}, a simple square or oblique lattice structure is often used, which misses the complexity in other three-dimensional lattices. Generally, not every family  of  lattice planes $(hkl)$ can produce  a constructive scattering.

Because  the density distribution $\rho({\bf r}) $   is periodic in crystals, one only needs to consider the particle distribution within one unit cell.  If each unit cell  has a  volume $V_{\rm cell}$ and  $m$ atoms,  then 
 \begin{align}
\hat{\rho}_{\bf q} =    \frac{N}{m}  \int\limits_{V_{\rm cell}}   d{\bf r}  \rho({\bf r})  e^{ i {\bf q} \cdot {\bf r}},
\end{align}
where $N/m$ is the number of unit cells in the $N$-particle system.
 In crystallography, it is customary to  define $\hat{\rho}_{\bf q}$ per unit cell as  the  {\em structure factor},
  \begin{align}
F_{\bf q}  & =   \int\limits_{V_{\rm cell}}   d{\bf r}  \rho({\bf r})  e^{ i {\bf q} \cdot {\bf r}}.
\end{align}
For crystals,  only  ${\bf q}$'s satisfying Bragg's law~(\ref{eq:bragg_q})  can possibly generate a large $\hat{\rho}_{\bf q} $ or $F_{\bf q} $.  Therefore,  we only need to consider ${\bf q}$'s  of the form
  \begin{align*}
 {\bf q}  = 2\pi {\bf d}^*_{hkl}= 2\pi (h {\bf a}^* + k {\bf b}^* + l  {\bf c}^*),
 \end{align*}
where ${\bf d}^*_{hkl}$ represents a family of lattice planes $(hkl)$ of spacing  $d_{hkl} = 1/ |{\bf d}^*_{hkl}|$ (Appendix~\ref{sec:lat}). The associated   structure factor can thus be   denoted as $F_{hkl}$ 
  \begin{align}
  \label{eq:Fhkl}
F_{hkl}   =  \int\limits_{V_{\rm cell}}  d{\bf r}  \rho(x,y,z)  e^{ 2\pi i  ( h x + k y + l z) }.
\end{align}
 Inversely,  the density distribution within each unit cell is 
  \begin{align}
\rho({\bf r})  = \frac{1}{V_{\rm cell}} \sum\limits_{hkl} F_{hkl}  e^{- 2 \pi i (h x + k y + l z) }.
\end{align}

For point-like scattering centers, $\rho({\bf r}) =  \sum\limits_{i=1}^m a_i  \delta ( {\bf r} -  {\bf r}_i ) $ and Eq.~(\ref{eq:Fhkl}) reduces to
 \begin{align}\label{eq:fhkl}
F_{hkl}    =   \sum\limits_{i=1}^m a_i  e^{ 2\pi i  ( h x_i + k y_i + l z_i) },
\end{align}
after substituting  Eq.~(\ref{eq:fpoint}) and following steps in Eq.~(\ref{eq:rhoq}), 
where $(x_i, y_i, z_i)$ are coordinates of  the $m$ atoms inside one unit cell and are expressed as fractions of   lattice vectors. The strength of $F_{hkl} $ by  planes $(hkl)$ is the vector sum of each term $a_i  e^{ 2\pi i  ( h x_i + k y_i + l z_i) }$ in Eq.~(\ref{eq:fhkl}), where  the phase angle $h x_i + k y_i + l z_i$ defines the direction of each vector.
 For typical crystal lattices of point-like atoms of the same type ($a_i = a$), $F_{hkl}$ can be easily computed.
 \begin{itemize}
 \item {\em Simple Cubic} (SC)

 $m=1$ and   $(x_1, y_1, z_1) = (0,0,0)$
  \begin{align}
F_{hkl}^{\rm SC}    =     a e^{ 2\pi i  ( h 0 + k  0 + l  0) } = a
\end{align}
for any $h,k,l$.
 \item {\em Body-Centered Cubic} (BCC)
 
 $m=2$, $(x_1, y_1, z_1) = (0,0,0)$ and $(x_2, y_2, z_2) = (1/2,1/2,1/2)$
  \begin{align}
 \begin{aligned}
F_{hkl}^{\rm BCC}   & =     a e^{ 2\pi i  ( h 0 + k  0 + l  0) }  + a e^{ 2\pi i  ( h \frac{1}{2} + k  \frac{1}{2} + l  \frac{1}{2} )}  \\
& = a +  a e^{ \pi i  ( h  + k   + l  ) }.
\end{aligned}
\end{align}

 \item {\em Face-Centered Cubic} (FCC)
 
 $m=4$, $(x_1, y_1, z_1) = (0,0,0)$,  $(x_2, y_2, z_2) = (1/2,1/2,0)$,  $(x_3, y_3, z_3) = (0, 1/2,1/2)$ and $(x_4, y_4, z_4) = (1/2,0, 1/2)$

 \begin{align}
 \begin{aligned}
F_{hkl}^{\rm FCC}   & =     a e^{ 2\pi i  ( h 0 + k  0 + l  0) }  + a e^{ 2\pi i  ( h \frac{1}{2} + k  \frac{1}{2} + l 0 )}  \\
& +    a e^{ 2\pi i  ( h 0 + k  \frac{1}{2} + l  \frac{1}{2} )} + a e^{ 2\pi i  ( h \frac{1}{2} + k 0 + l  \frac{1}{2} )}  \\
& = a +  a e^{ \pi i  ( h  + k      ) }  + a e^{ \pi i  (  k   + l     ) }  + a e^{ \pi i  ( h  + l       ) } .
\end{aligned}
\end{align}

 \end{itemize}
The $F_{hkl}$  of  BCC and FCC lattices   completely vanishes for certain $h,k,l$. The resulting   reflection Miller indices should successively be $(110)$, $(200)$, $(211)$, $(220)$, $(310)$, $(222)$ $\cdots$  for BCC,  and $(111)$, $(200)$, $(220)$, $(311)$, $(222)$, $(400)$ $\cdots$ for FCC crystals.

\subsection{Finite-Size Crystals and Bragg Peak Broadening}

When Bragg's law is satisfied by wavelength $\lambda$ at an incident angle $\theta$, a small deviation $\delta \theta$ from $\theta$ only slightly changes the path difference between two rays reflected by a pair of neighboring planes (of spacing $d_{hkl}$), which  still add constructively.  If we consider two reflection planes that are $2d_{hkl}$, $3d_{hkl}$, $\cdots$, apart, the change in path difference due to $\delta \theta$ accumulates, and at large enough spacing, becomes $\lambda/2$ such that the two rays completely cancel. For a beam reflected by a  crystallographic plane in {\em  large} crystal samples, it is always possible to find another remote  plane whose reflected beam interferes  destructively,  even for very small $\delta \theta$. Therefore, when other broadening effects are excluded, diffraction signals  in large samples at fixed $\lambda$, if there are any,  should in principle be of infinitely small size (in terms of the range of $\theta$).
\begin{figure}
\centering
\includegraphics[width=0.45\textwidth]{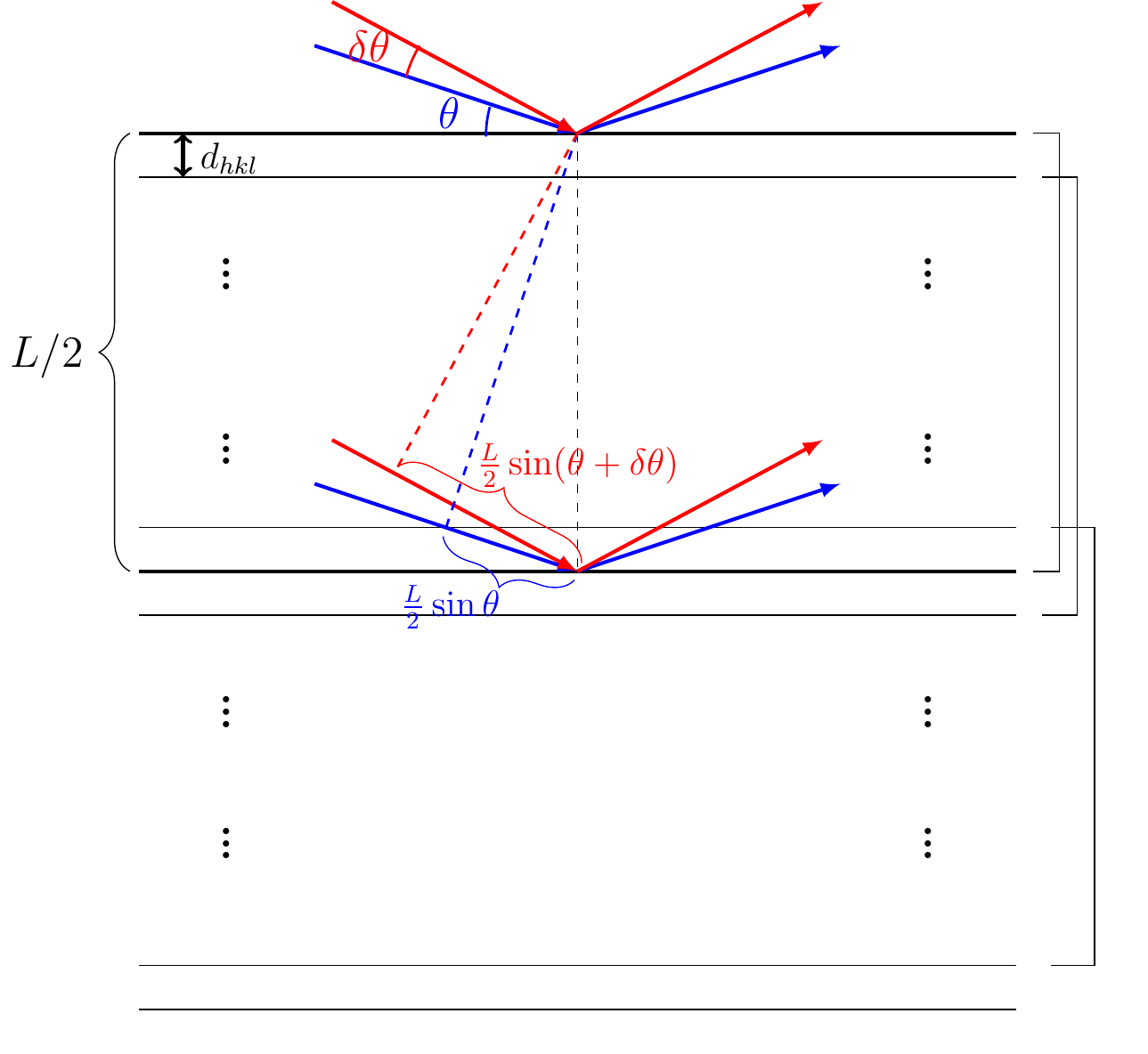}
\caption{Illustration of the Scherrer equation. In a finite-size crystal of thickness $L$, the path difference between two rays reflected by a pair of planes that are $L/2$ apart sets the limit of the signal broadening $2\delta\theta$. }
\label{fig:sch}
\end{figure} 

For {\em small} crystal samples,  it is possible that the  change in  path difference  due to $\delta \theta$  is  much less than $\lambda/2$ such that Bragg's law is still  approximately satisfied at $\theta + \delta \theta$ and the diffraction signal is broadened by an amount $\sim \delta \theta$. The quantitative relationship between the broadening $2\delta \theta$ of the signal and  the linear dimension $L$ of a finite-size crystal can be found by considering all pairs of planes that are $L/2$ apart. When $\theta$ changes to $\theta + \delta \theta$, the path difference  for such a pair of planes increases by  $2  \frac{L}{2} \left [   \sin(\theta + \delta \theta)-  \sin\theta \right] = L\cos\theta \delta\theta $ (Fig.~\ref{fig:sch}). The diffraction signal broadens  until destructive interference occurs at
$\frac{\lambda}{2} =  L \cos\theta \delta\theta$,
which gives the {\em Scherrer equation}
  \begin{align}
 \begin{aligned}
2 \delta \theta = \frac{\lambda}{ L \cos \theta} = \frac{2\tan \theta}{L/d_{hkl}}.
\end{aligned}
\end{align}
Thus diffraction signals tend to be larger in smaller systems. 

\section{Liquid-State Theory}

\label{sec:liquid}
According to liquid-state theory, a {\em static structure factor }   $S({\bf q}) $ can be used to address short-range order~\cite{thomas1941,rahman1962} and the  glass transition~\cite{janssen2018} in amorphous/liquid samples~\cite{fischer2006} and more generally in  nano-structured or other structurally disordered systems~\cite{Billinge2019}. In  an $N$-particle system,  it  is defined as
 \begin{align}
S({\bf q}) =  \frac{1}{\sum\limits_{i=1}^N   \hat{f}_i^2( {\bf q})  }   \left \langle \hat{\rho}_{\bf q} \hat{\rho}_{\bf -q}  \right \rangle =   \frac{1}{\sum\limits_{i=1}^N   \hat{f}_i^2( {\bf q})  }   I({\bf q}), 
\end{align}
where the ensemble average $\langle \cdots \rangle$ is usually taken over  configurations at thermal equilibrium~\cite{hansen:2013}.   Practically, this ensemble average results from a sum over all the different coherence  volumes in the sample, after being Fourier transformed, giving  a real-space representation of the sample's ensemble-averaged instantaneous local structure.
 
If  scattering  centers are point-like, i.e.  $ \rho({\bf r})  =  \sum\limits_{i=1}^N   a_i  \delta( {\bf r} - {\bf r}_i) $,   then $\hat{\rho}_{\bf q}   = \sum\limits_{i=1}^N   a_ i e^{ i {\bf q} \cdot {\bf r}_i} $ and 
\begin{align}
S({\bf q})  = \frac{1}{\sum\limits_{i=1}^N  a_i^2 }   \left \langle     \left |      \sum\limits_{i=1}^N a_i   \cos({\bf q} \cdot {\bf r}_i ) \right|^2  +  \left |      \sum\limits_{i=1}^N  a_i \sin({\bf q} \cdot {\bf r}_i ) \right|^2  \right\rangle.
\end{align}
For monodisperse systems ($a_i$ is the same for all particles), $S(0) = N$.

In the case of $a_i = 1$, S({\bf q}) is related to 
the radial distribution function $g({\bf r})$ or the pair correlation function $h({\bf r}) = g({\bf r}) -1$ by
\begin{align}
\begin{aligned}
S({\bf q})   &=  1 + \rho_0 \int  d {\bf r}   (g({\bf r}) -1) e^{i  {\bf q} \cdot {\bf r}_i } + \rho_0 \int  d {\bf r}   e^{i  {\bf q} \cdot {\bf r}_i }  \\
 &=  1 + \rho_0 \int  d {\bf r}   (g({\bf r}) -1) e^{i  {\bf q} \cdot {\bf r}_i } + \rho_0 (2\pi)^3 \delta_D({\bf q})\\
  &=  1 + \rho_0 \int\limits_V  d {\bf r}   (g({\bf r}) -1) e^{i  {\bf q} \cdot {\bf r}_i }  + \rho_0 V \delta_{ {\bf q} , 0} ~~({\rm finite}~ V) \\
&=  1 + \rho_0 \int\limits_V d {\bf r}   h({\bf r})  e^{i  {\bf q} \cdot {\bf r}_i }  + N \delta_{ {\bf q} , 0}  \\
&=  1 + \rho_0  \hat{h}_{{\bf q}}+ N \delta_{ {\bf q} , 0}
\end{aligned}
\end{align}
where the global number density $\rho_0 = N/V$  and the Fourier transform $ \hat{h}_{{\bf q}}  =\int\limits_V  d {\bf r}   h({\bf r})  e^{i  {\bf q} \cdot {\bf r}_i } $.
Note that $S({\bf q})  $ is singular or discontinuous at ${\bf q} = 0$, i.e. $\lim\limits_{{\bf q} \to 0}  S({\bf q}) \ne S(0) = N$. Correspondingly, 
$\lim\limits_{{\bf q} \to 0}   \hat{h}_{{\bf q}}  \ne  \hat{h}_{0} = -1/\rho_0$.

The radial distribution function can be obtained from the structure factor by the inverse Fourier transform
\begin{align}
g({\bf r})  = 1 + \frac{1}{(2\pi)^3} \int\limits_{{\bf q} \to 0} d {\bf q} \frac{S({\bf q})  - 1}{\rho_0} e^{ - i  {\bf q} \cdot {\bf r}_i }, 
\end{align}
where the value  $\lim\limits_{{\bf q} \to 0}  S({\bf q}) $ should be used  at ${\bf q} = 0$ in the integration.
When the system's  structure is isotropic over the sample volume, i.e. $g({\bf r})  = g(r)  $, more convenient relationships can be derived~\cite{keen2001} 
\begin{align}
\label{eq:sq_gr}
S(q)  = 1 +  4\pi \rho_0  \int_{0^+}^{\infty} dr (g(r) - 1)  r^2 \frac{\sin(qr)}{qr}, \\
g(r)  = 1 + \frac{1}{2\pi^2} \int_{0^+}^{\infty} d q \frac{S(q)  - 1}{\rho_0}  q^2 \frac{\sin(qr)}{qr}, 
\end{align}
where $\lim\limits_{x \to 0} \frac{ \sin x}{ x} = 1$ should be used in the integration.  The limit value of  $S(q)$ as $q$ approaches zero is related to the isothermal compressibility $\kappa$ by~\cite{barrat:2003}
\begin{align}
\lim\limits_{q \to 0}  S(q) = \rho_0 k_B T \kappa.
\end{align}

\section{Experimental Setups in  Photography}

\label{sec:exp}
In this section,  we discuss some technical details about  photography methods, which collects signals of $I({\bf q}) = I(X,Y)$ on a two-dimensional film with coordinates $(X,Y)$. Three popular  experimental setups are often used as described below, which map ${\bf q}$ onto $(X,Y)$ differently.

\subsection{Back-reflection and Transmission Methods}

\begin{figure*}
\centering
\includegraphics[width=0.45\textwidth]{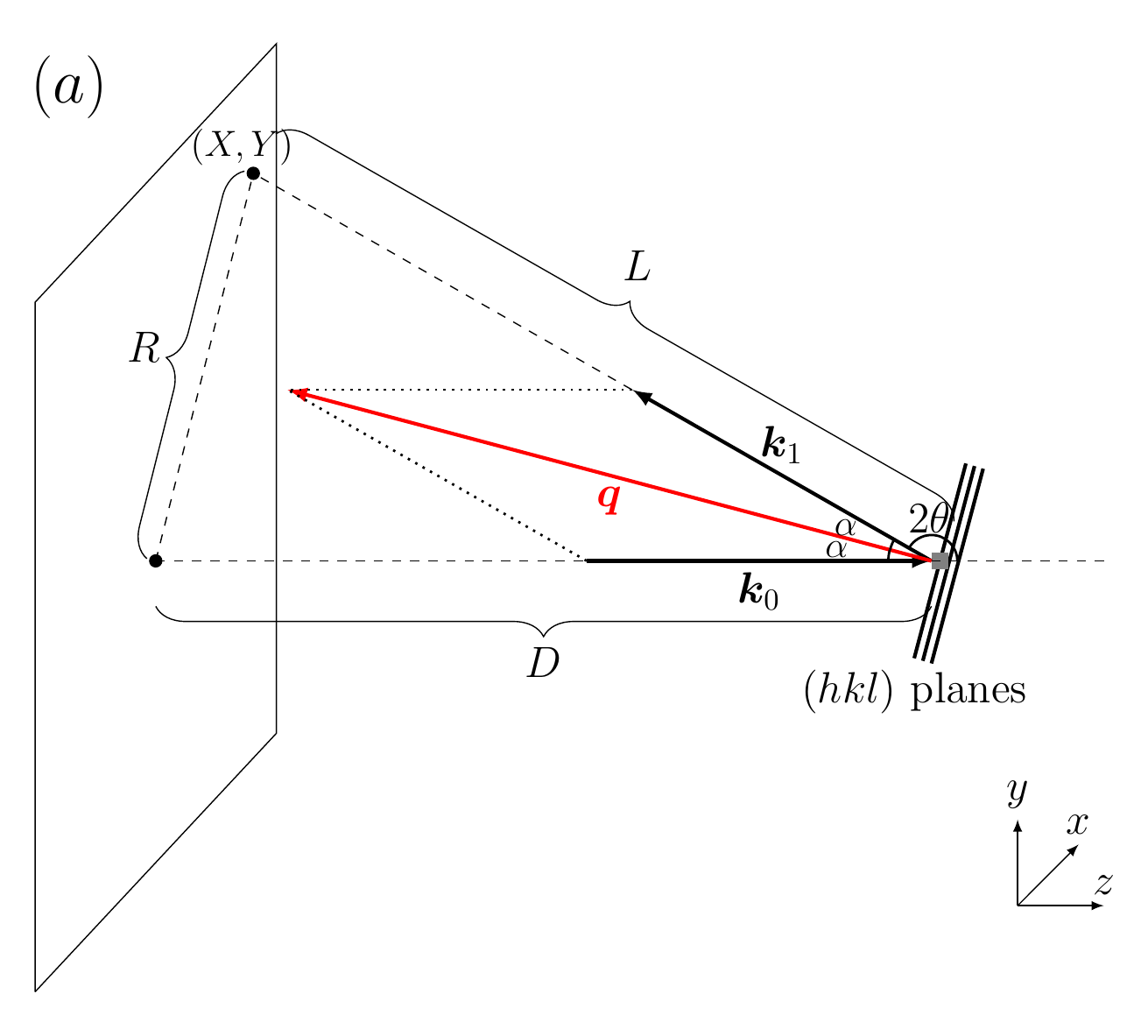}
\includegraphics[width=0.35\textwidth]{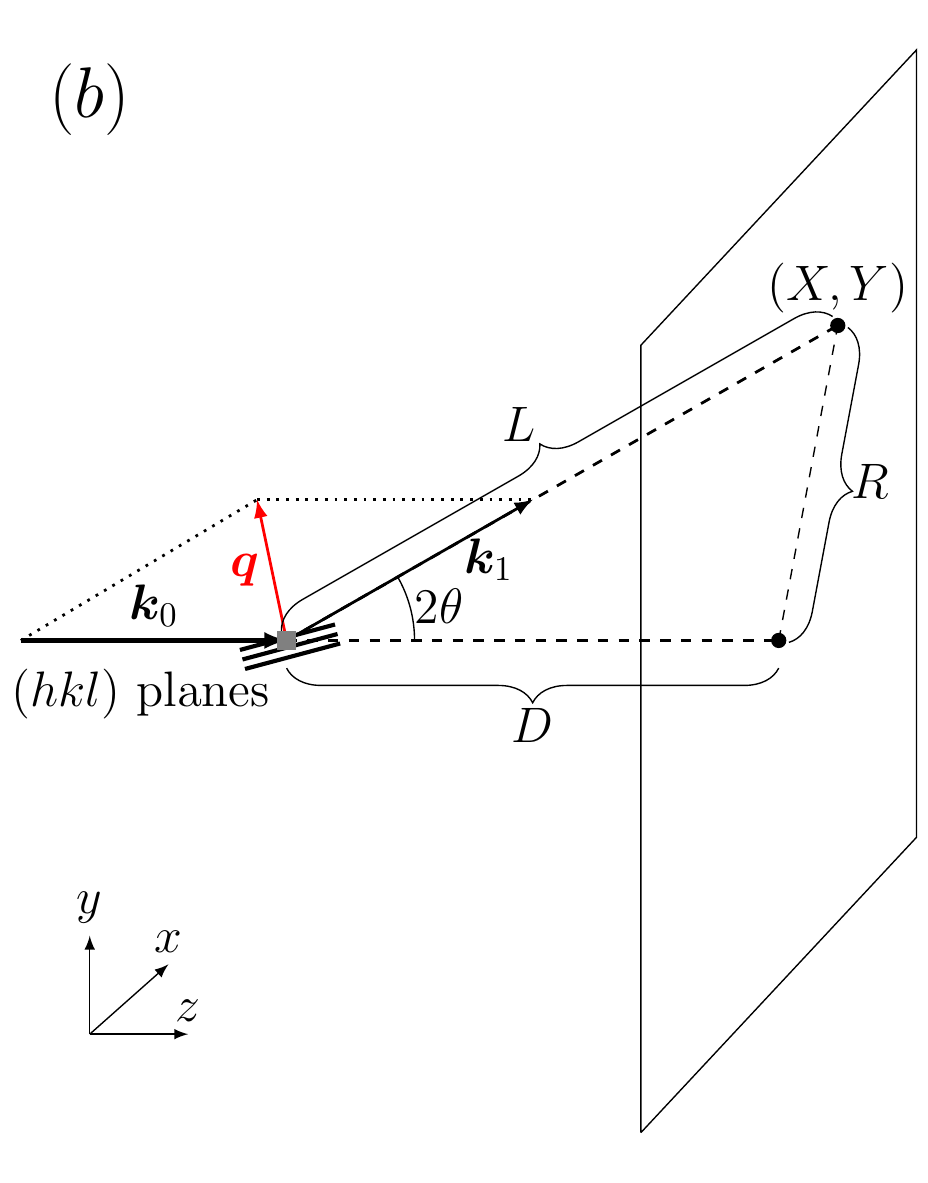}
\caption{Illustration of  (a) back-reflection and  (b)  transmission methods. The scattering vector ${\bf q}$ resulted from crystallographic planes $(hkl)$ maps onto 2D coordinates $(X,Y)$ on the film, which is placed at a distance $D$ from the sample. The incident beam is along the $z$ axis. }
\label{fig:plane}
\end{figure*}

In  back-reflection and transmission methods, the recording film is a rectangular plane, which is placed either before (back-reflection) or after (transmission) the sample  as shown in Fig.~\ref{fig:plane}. In both methods, it can be seen that the ratio $q_x/q_y$ equals $X/Y$. If the incident wave number is $|{\bf k}_0| = 2\pi/\lambda$, then
\begin{align}
\label{eq:qxqy}
(q_x, q_y) = \frac{2\pi}{\lambda} \left(\frac{X}{L},\frac{Y}{L} \right),
\end{align}
where $L^2 = R^2 +D^2$ and  $R^2 = X^2 + Y^2 $. The difference remains in the $z$ component $q_z$.

In the back-reflection method, because $\alpha = \frac{\pi}{2} - \theta  $ satisfies   $\cos (2\alpha) = D/L $, it follows that 
\begin{align*}
 \begin{aligned}
q_z &= - 2  \frac{2\pi}{\lambda}  \cos \alpha \cos \alpha \\
& = -  \frac{2\pi}{\lambda}  [1 + \cos( 2\alpha) ] \\
& =   -  \frac{2\pi}{\lambda} (1 + D/L).
\end{aligned} 
\end{align*}
Therefore,
\begin{align}
 (q_x, q_y, q_z) = \frac{2\pi}{\lambda} \left(\frac{X}{L},\frac{Y}{L}, - \frac{D+L}{L}  \right) \mbox{ (back-reflection)}  
\end{align}
In contrast, one can show that, in the transmission method,
\begin{align}
\label{eq:q_trans}
(q_x, q_y, q_z) = \frac{2\pi}{\lambda} \left(\frac{X}{L},\frac{Y}{L}, - \frac{L-D}{L} \right)  \mbox{ (transmission).}  
\end{align}

\subsection{Cylindrical Method}

\begin{figure*}
\includegraphics[width=0.35\textwidth]{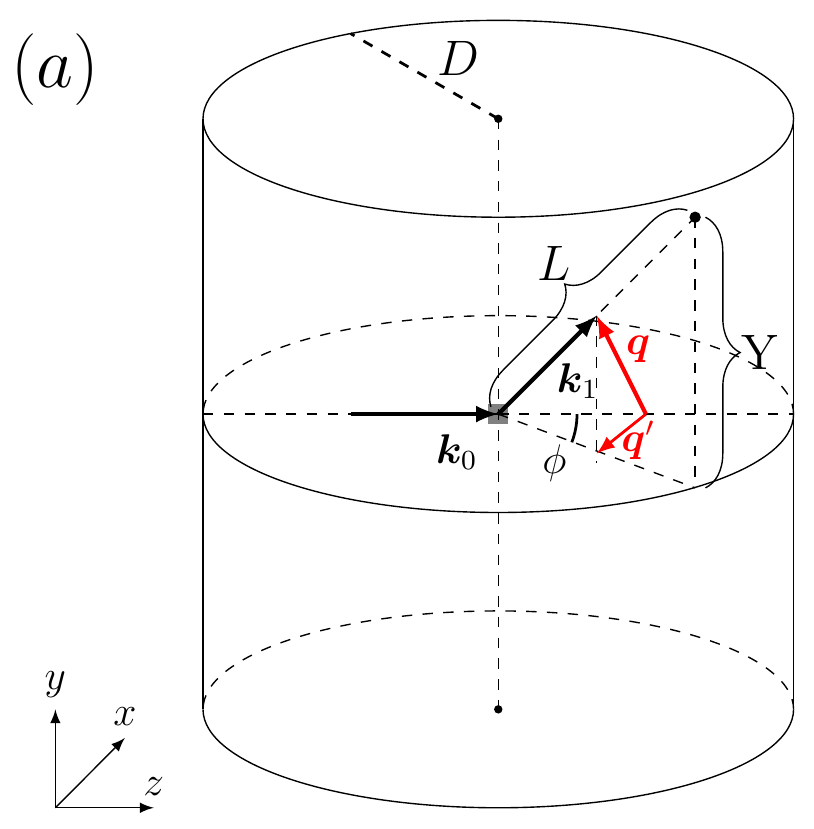}
\includegraphics[width=0.45\textwidth]{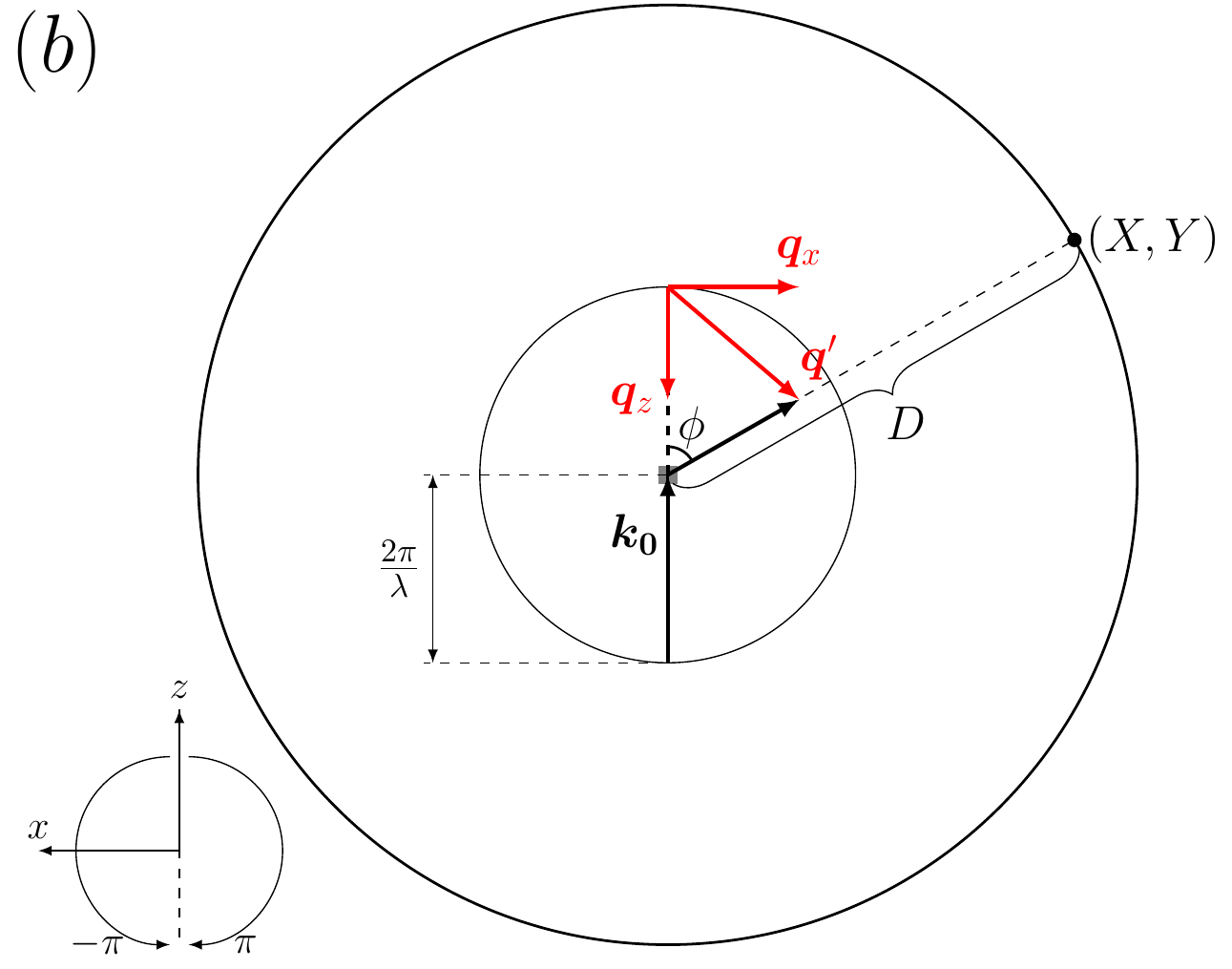}
\caption{ Illustration of cylindrical method from (a) the side view and (b) the  top view. A cylindrical film is placed at a radius $D$ around the sample. }
\label{fig:cylinder}
\end{figure*}

Compared with above two setups, the cylindrical method collects signals from all azimuthal angles $\phi$ and is thus more informative (Fig.~\ref{fig:cylinder}).  In fact,  a cylindrical film, which better preserves the natural shape of scattering spots,  can be considered as the sum of an infinitely wide back-reflection film and an infinitely wide transmission film,  on which scattering patterns farther away from the  film center are more distorted.   

To map ${\bf q}$ onto the film, one can unfold the cylinder into a plane with coordinates $(X,Y) = (D\sin \phi, Y)$ with the azimuthal angle $\phi\in(-\pi, \pi)$. The relationship is 
\begin{align}
(q_x, q_y, q_z) = \frac{2\pi}{\lambda} \left(\frac{D \sin \phi}{L},\frac{Y}{L},  \frac{D\cos \phi - D}{L} \right).
\end{align}

\section{Photography of Single Crystalline Samples}

\label{sec:photo}
The illustration of Bragg's law using Ewald's sphere suggests two ways to make reciprocal lattice points fall on the sphere   to generate constructive  scattering signals from specific crystallographic planes. One is to tune the wavelength and the other is to change the orientation of the sample. These correspond to two experimental strategies in designing  photography methods for single crystals -- the Laue method and the (monochromatic) rotation method.

\subsection{Varying Wavelength at Fixed Angle -- Laue Method}

In the Laue method, one fixes  the orientation of the sample (thus the angle $\theta$ in Bragg's law) and changes the wavelength of incident beam over a certain range $\lambda \in [\lambda_{\min}, \lambda_{\max}]$, which is thus called ``white color''.

For each pixel $(X,Y)$ on the film, the scattering intensity is thus the sum of  contributions  from all wavelengths, or equivalently, all parallel scattering vectors ${\bf q}$, which can be formally written as
\begin{align}
\label{eq:I_laue}
\begin{array}{ll}
&\bar{I}(X, Y) =  \sum\limits_{\bf q}  I({\bf q}) \\
& =   \sum\limits_{\bf q}  \left[ \left |      \sum\limits_{i=1}^N \hat{f}_i( {\bf q})   \cos({\bf q} \cdot {\bf r}_i ) \right|^2  +  \left |      \sum\limits_{i=1}^N  \hat{f}_i( {\bf q})  \sin({\bf q} \cdot {\bf r}_i ) \right|^2   \right].
\end{array}
\end{align}
This type of general equation, which computes scattering intensity from all atoms in the sample, reduces to the simple summation over atoms in the unit cell for ideal crystals, as explained in Section~\ref{sec:Fhkl}.
\begin{figure}
\includegraphics[width=0.5\textwidth]{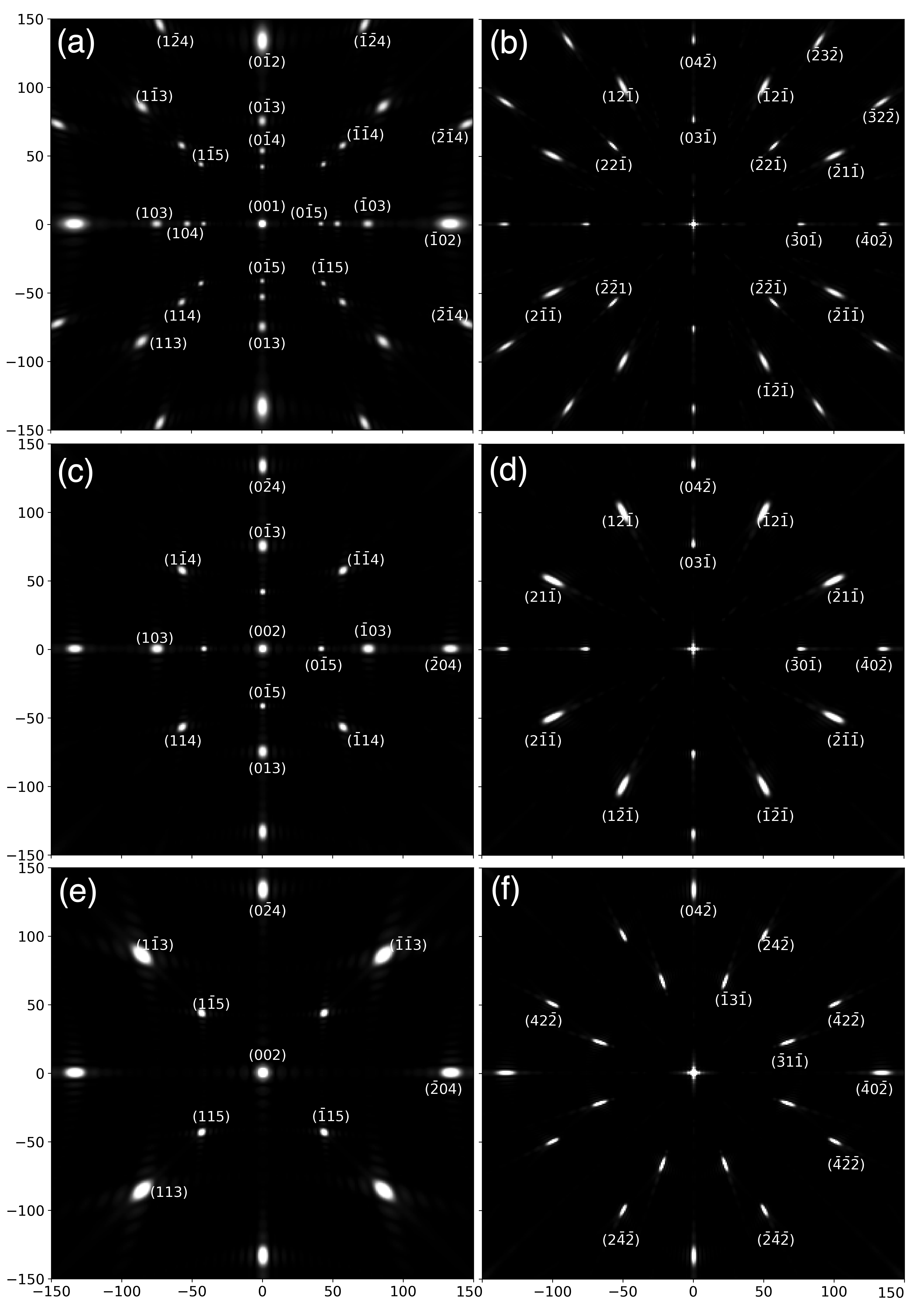}
 \caption{ Back-reflection (left column) and  transmission (right column) photography of SC (a-b)  ($N = 3375$), BCC (c-d)  ($N = 4394$) and FCC (e-f)   ($N = 5324$) crystals. The range of  wavelength   $\lambda$ is 0.35-1.0$\sigma$ for SC back-reflection, 0.199-0.35$\sigma$  for SC transmission, 0.4-1.2$\sigma$ for BCC back-reflection, 0.23-0.4$\sigma$  for BCC transmission,  0.5-1.42$\sigma$ for FCC back-reflection and 0.23-0.49$\sigma$  for FCC transmission. The $(X,Y)$ coordinate range $[-150,150]$ is now set by the grid resolution of the computer code, which can be mapped onto the real length unit on a physical film.}
\label{fig:Ixy_laue}
\end{figure}

We demonstrate the photography results  using  perfect SC, BCC and FCC samples (Fig.~\ref{fig:Ixy_laue}). The incident beam is  along the $[001]$ direction and the nearest neighbor distance $\sigma$ is set as the unit of length.  The  code to compute $\bar{I}(X, Y) $  numerically  implementing Eq.~(\ref{eq:I_laue})  is provided online. The value of $D$ can be chosen arbitrarily, with all other lengths calculated accordingly,  because it only leads to a scaling of the photograph.  Here, we set $D = 100\sigma$  for numerical convenience.   If the  total number of pixels on the film is  $N_{XY}$ and the number of  wavelengths scanned is $N_{\lambda}$, then the computational complexity  using Eq.~(\ref{eq:I_laue}) is ${\mathcal O}(N_{XY}N_{\lambda}N)$.

\subsection{Varying Angle using Fixed Wavelength -- Rotation Method}

\begin{figure*}
\centering
\includegraphics[width=0.95\textwidth]{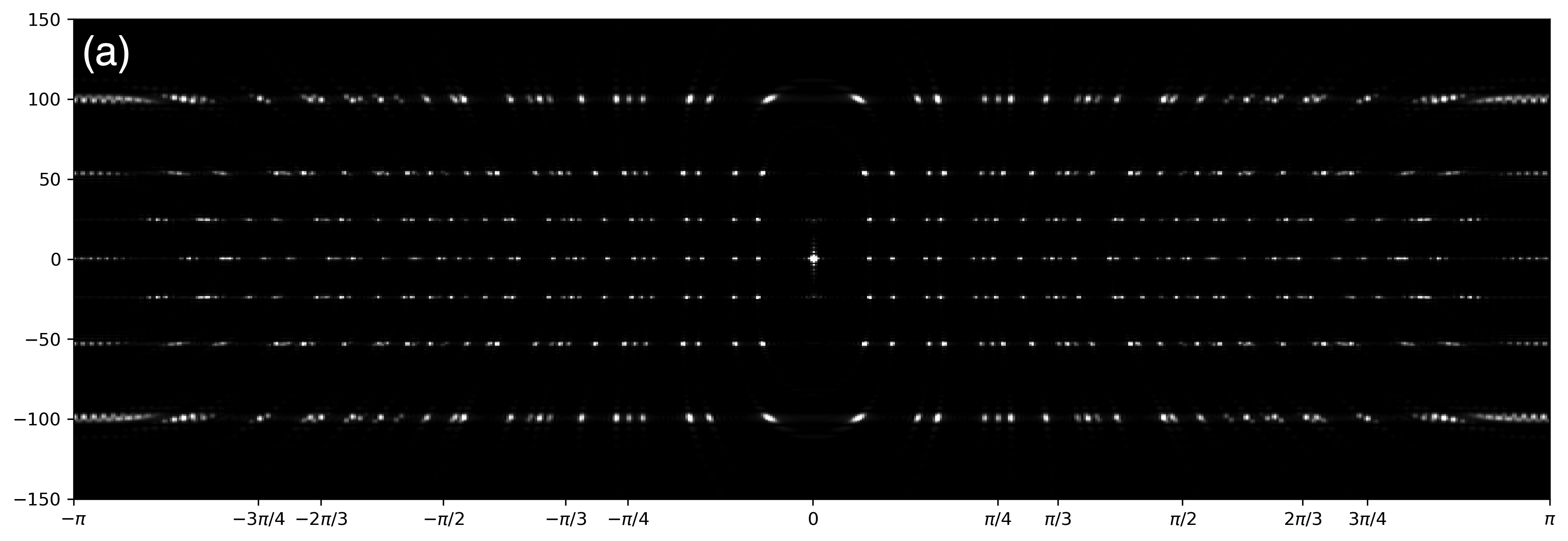}
\includegraphics[width=0.95\textwidth]{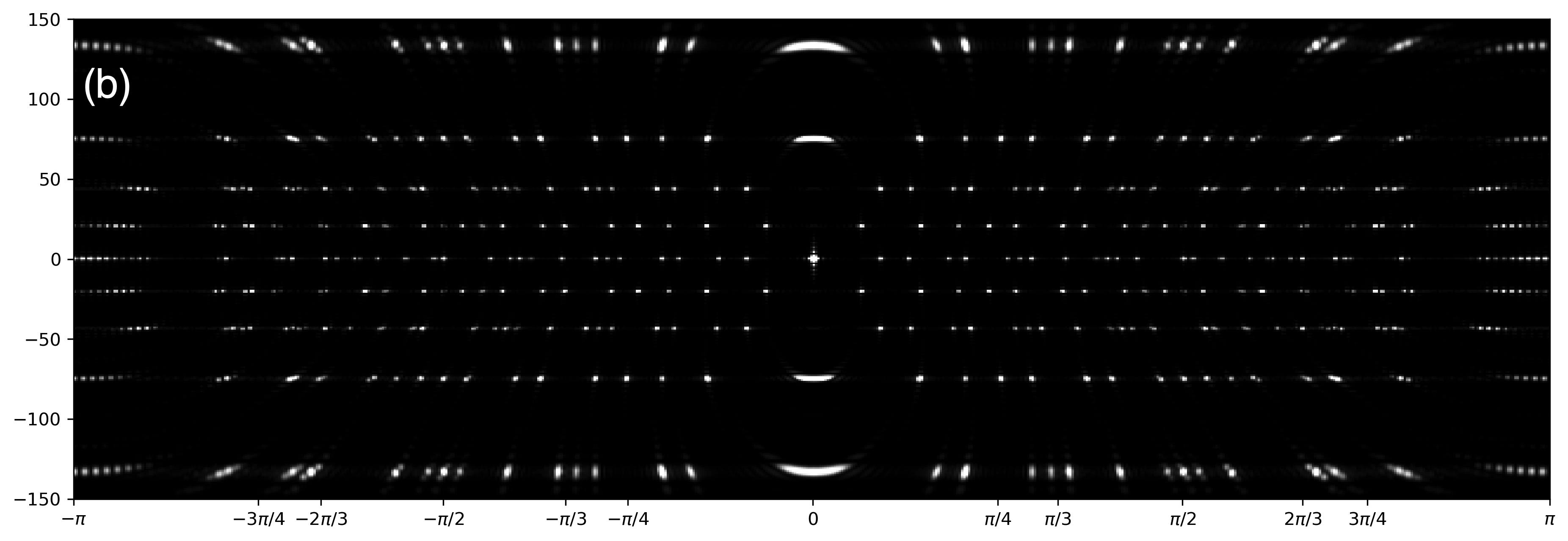}
\includegraphics[width=0.95\textwidth]{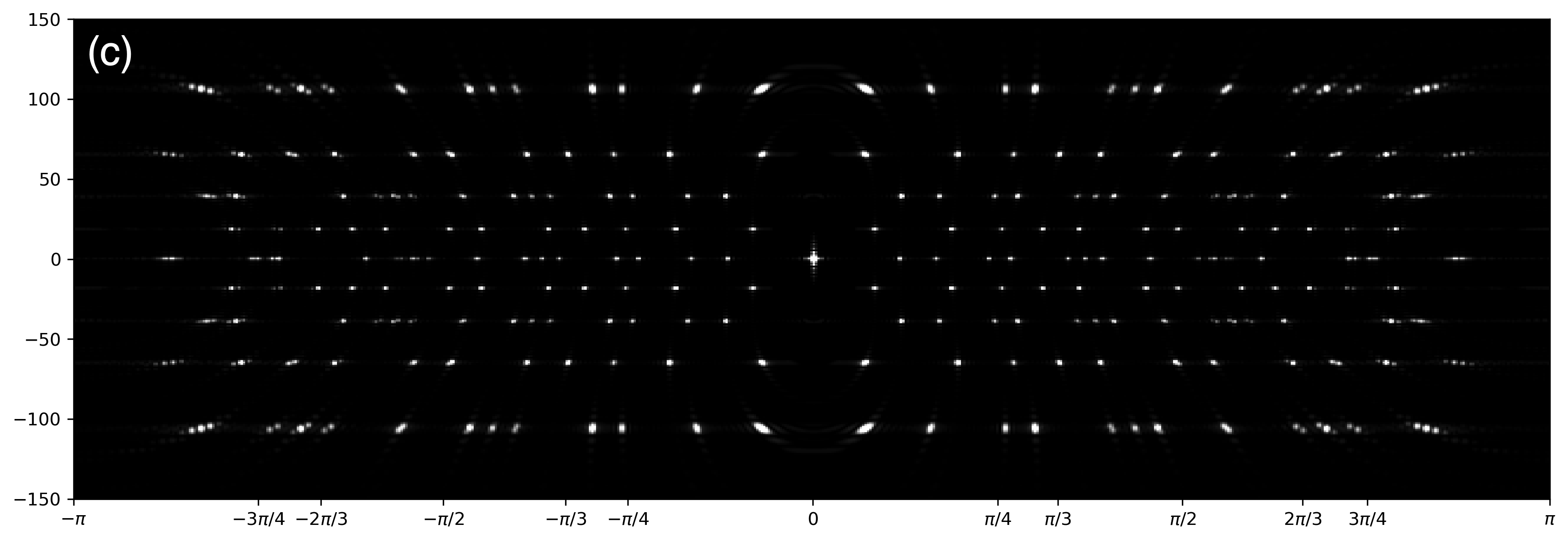}
\caption{Rotation photography of SC (a) ($N = 3375$ and $\lambda = 0.235\sigma$), BCC (b) ($N = 4394$ and $\lambda = 0.231\sigma$) and FCC (c)   ($N = 5324$ and $\lambda = 0.257\sigma$) crystals.}
\label{fig:Ixy_rot}
\end{figure*}

We use the conventional setup -- cylindrical film -- to explain the rotation method for the same SC, BCC and FCC crystalline samples as above (Fig.~\ref{fig:Ixy_rot}). The wavelength $\lambda$ of the incident beam is fixed in this method and the sample placed at the central axis of cylinder is rotated by a certain angle to probe   possible orientations and scattering angle $2\theta$ for given crystallographic planes. A full circle of $2\pi$ rotation is only necessary for noncentrosymmetric crystals containing elements that exhibit anomalous dispersion; a rotation of $2\pi$ is sufficient for centrosymmetric crystals.

The scattering intensity at coordinates $(X,Y)$ is then 
\begin{align}
\label{eq:I_rot}
\begin{array}{ll}
& \bar{I}(X, Y) =  \\
 & \sum\limits_{\Omega}   \left[ \left |      \sum\limits_{i=1}^N \hat{f}_i( {\bf q})   \cos({\bf q} \cdot {\bf r}_i (\Omega) ) \right|^2  +  \left |      \sum\limits_{i=1}^N  \hat{f}_i( {\bf q})  \sin({\bf q} \cdot {\bf r}_i(\Omega) ) \right|^2   \right]\\
\end{array}
\end{align}
where $\Omega$  represents the orientation of the sample due to rotation. For a given sample, we apply a rotation matrix   about its $y$-axis to transform   particle coordinates into new values. The  accumulated signal  $\bar{I}(X, Y) $  on the cylinder is then unfolded onto a rectangle. If a total number $N_{\Omega}$ of rotation angles within $(0, 2\pi)$ are scanned, the computational complexity to implement Eq.~(\ref{eq:I_rot}) to produce results on $N_{XY}$ pixels is then   ${\mathcal O}(N_{XY}N_{\Omega}N)$.

\subsection{Broadening due to Finite-size Effect}

So far we assume that either varying wavelength or varying sample orientation is needed to satisfy Bragg's law and produce nonvanishing scattering signals on  the photograph. However, this is only true for infinitely large systems.    In our small samples with $N\sim 10^3$ particles, signal broadening allows us to observe certain scattering patterns, even when the wavelength $\lambda$ is fixed at one appropriate value.

\begin{figure}
\centering
\includegraphics[width=0.45\textwidth]{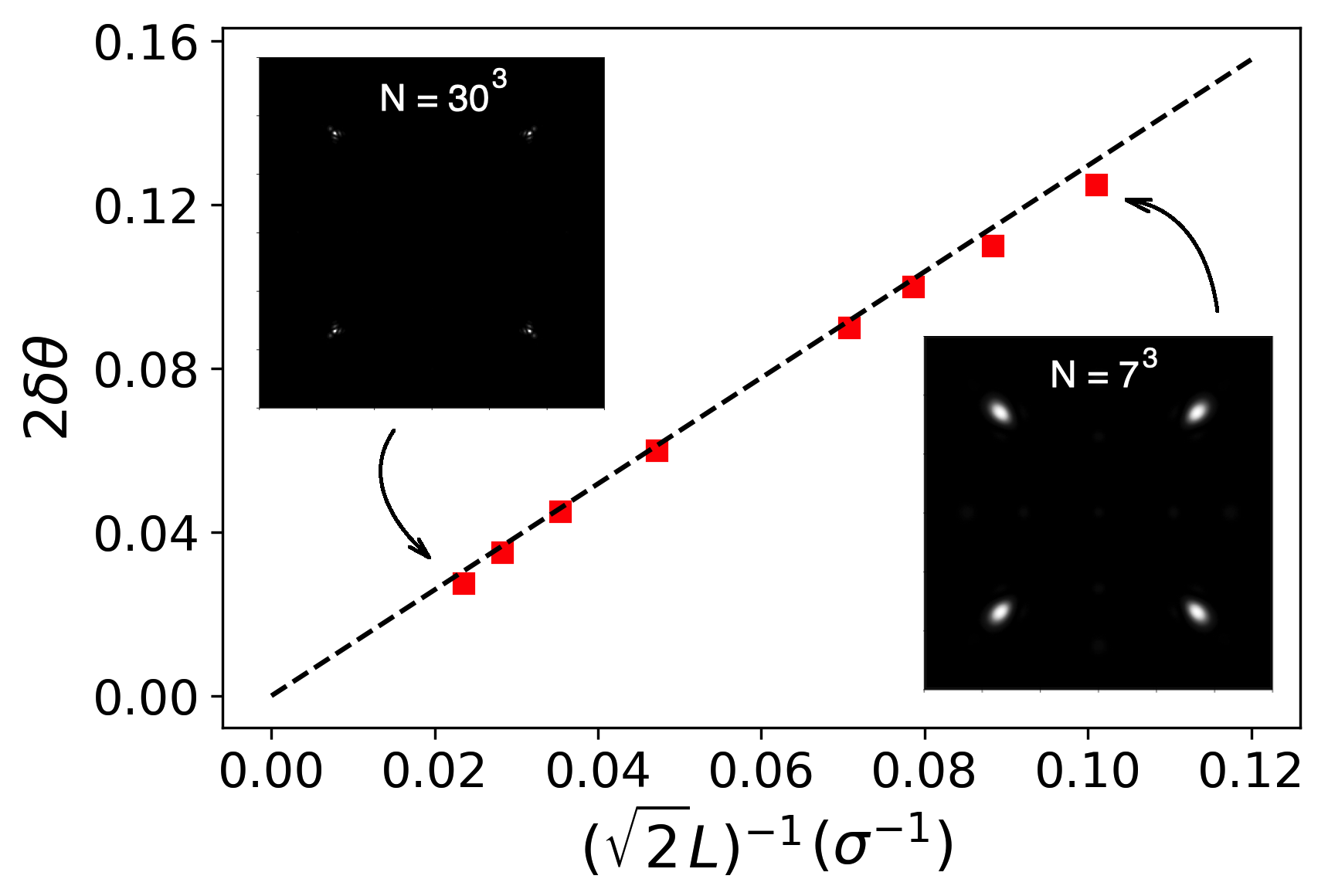}
\caption{System size effect on scattering size in back-reflection of  SC samples with fixed wavelength $\lambda = 0.55\sigma$.  The scattering angle $2\theta$ for these four spots can be computed from $\tan (\pi - 2\theta) =85\sqrt{2}/100$.  After scaling $L$ by a factor of $\sqrt{2} $, the data (red squares) agree with theoretical slope (dashed line) from Scherrer equation $\lambda/\cos\theta$. }
\label{fig:Ixy_size}
\end{figure}

For example, in the previously mentioned SC crystals, we can see four scattering spots in the back-reflection method at fixed wavelength $\lambda = 0.55 \sigma$, which correspond to the $(113)$ planes and equivalents (Fig.~\ref{fig:Ixy_size}). When the system size is varied from $N=7^3$ to $30^{3}$, the size of each spot decreases. It can be confirmed that the relationship between box size $L = \sqrt[3]{N}$ and spot size $2\delta\theta$ roughly satisfies the Scherrer equation $2\delta\theta \propto 1/L$.  An empirically scaling factor $\sqrt{2}$ is needed on $L$ to estimate the actual dimension of the sample perpendicular to $(113)$ planes and to agree with the theoretical slope $\lambda/\cos\theta$.

\subsection{DNA Double Helix}

\begin{figure*}
\centering
\includegraphics[width=0.95\textwidth]{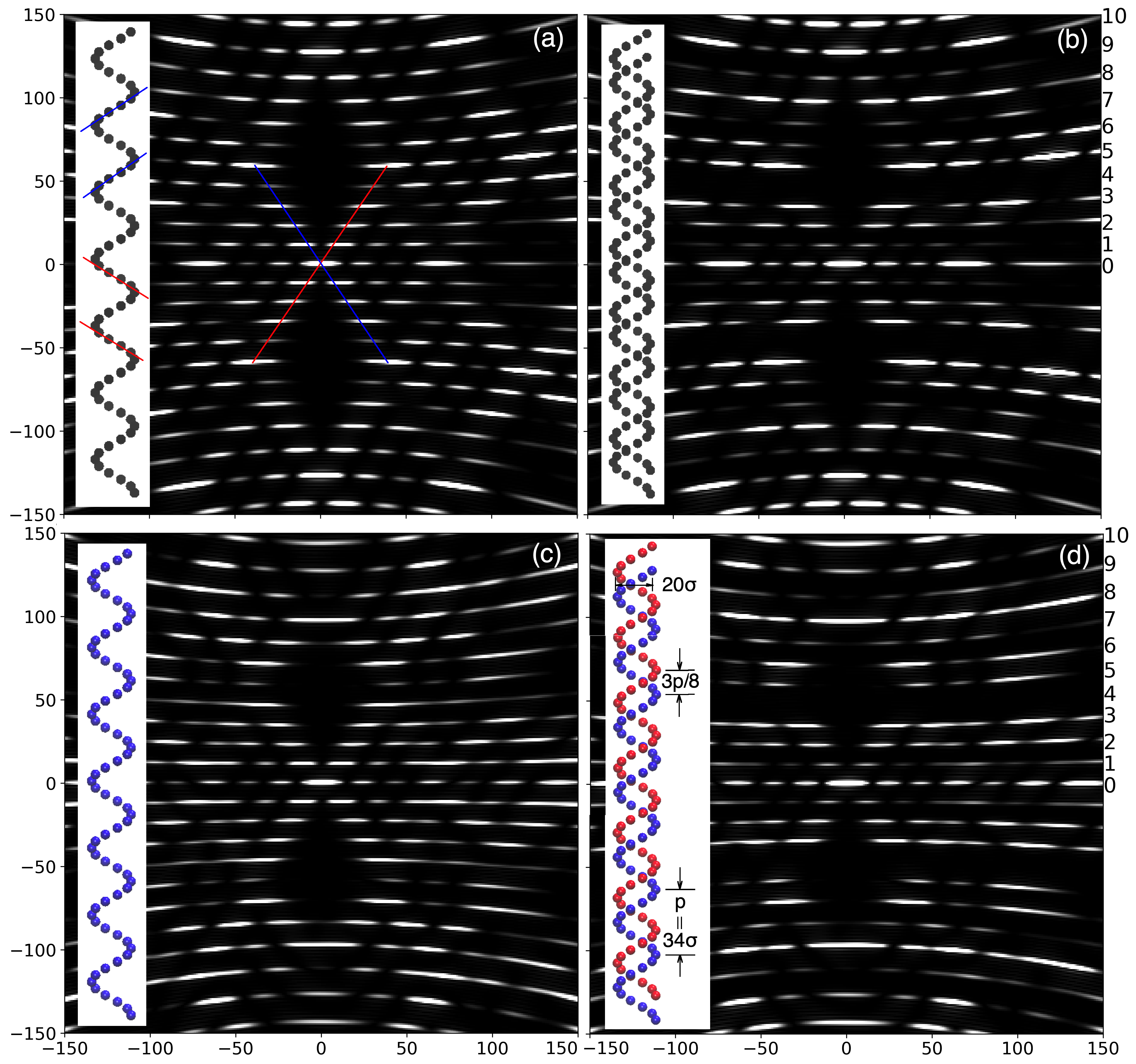}
\caption{Transmission photography of single-strand (a,c) and double-strand (b,d) 2D sinusoidal waves (a-b) and 3D DNA helices (c-d) using $\lambda = 1.54\sigma$. Each helix is made of a backbone of $N = 70$ particles with a pitch of $p = 34\sigma$ and a diameter of $20\sigma$. There are 10 particles per pitch. The two helices in the double strand structure are offset by $3/8$ pitch. Particle size in  insets is set as $5\sigma$ to enhance visibility.}
\label{fig:Ixy_dna}
\end{figure*}

One of the most successful and famous applications of scattering methods is the determination of the
DNA structure, whose X-ray photography shows a characteristic ``X''-shape pattern with horizontal stripes~\cite{franklin1953}. The form of the pattern can be understood analytically by   diffraction from the 2D projected sinusoidal waves of the single or double helix~\cite{kittel1968,thompson2018}. Here we produce the transmission photography of a single model DNA  fiber with only backbone particles. Each helix has $N=70$ particles with $10$ particles per turn (pitch). The parameters of the right-handed B-DNA, $34\sigma$  for pitch and $20\sigma$ for helix diameter, are used (Fig.~\ref{fig:Ixy_dna}). The unit of length $\sigma$ can be mapped onto the real length unit \AA.

All the four photographs, with the fiber being single  or double strand, 2D projected or 3D stereoscopic,  have an ``X''-shape pattern at the center and are made of horizontal broken stripes (Fig.~\ref{fig:Ixy_dna}).  The two branches of the ``X'' pattern   can be viewed as scattering signals from the two series of parallel particles on the sinusoidal wave (Fig.~\ref{fig:Ixy_dna}a). Because each pitch of the helix has 10 particles, the pattern has a vertical period of 10 stripes~\cite{kittel1968}. The  brightness and darkness along each horizontal stripe depends sensitively on the relative position between different particles~\cite{lucas2005}. For example, the level 4 stripe disappears when two double strands with a phase difference of $3/8$ pitch are present. The bright level 8 signal of 3D samples at $X=0$ is missing for   2D structures.

\section{Scattering Vector ${\bf q}$ in Intensity Scanning}

\label{sec:q}
In this section,  we discuss the choice of scattering vector ${\bf q}$ in  the case of  disordered or partially ordered samples that are spatially isotropic or approximately isotropic. When samples are  isotropic, the scattering intensity $I({\bf q})$ or its normalized version, structure factor $S({\bf q})$,  only depends on the magnitude $q$ of the scattering vector, and thus does not  generate isolated  spotty signals as in photography of ordered samples.  The photography $I(X, Y)$, often of less interest in this context, should ideally exhibit concentric circular patterns.  The intensity scanning $I(q)$ or $S(q)$ as a function of $q$ is the primary method used for isotropic samples.

\subsection{Vector ${\bf q}$ along a Single Direction  to  Represent Magnitude $q$ in  Isotropic Systems}

 In an experiment, one can vary $q$ by observing signals at continuously changing scattering angle $2\theta$ using a fixed  incident wavelength $\lambda$.  Because experimental samples are generally large enough, a well-averaged  scattering signal can be detected along {\em one} particular direction at  $2\theta$, as in the powder method with a diffractometer. 

For example, consider a polycrystal with $M$ randomly oriented crystalline grains (domains), each of $N$ particles. The scattering intensity at ${\bf q}$ computed from Eq.~(\ref{eq:Iq}), assuming $\hat{f}_i( {\bf q})  = 1$,  is
\begin{align}
\label{eq:Iq_exp}
I({\bf q})   = \left |   \sum\limits_{n=1}^M      \sum\limits_{i=1}^N     \cos({\bf q} \cdot   {\bf r}_{n,i}   ) \right|^2  +  \left |    \sum\limits_{n=1}^M      \sum\limits_{i=1}^N   \sin({\bf q} \cdot {\bf r}_{n,i} ) \right|^2, 
\end{align}
where ${\bf r}_{n,i} $ is the position vector of particle $i$ in grain $n$. If $M$ is large and crystalline grains are  uniformly oriented in all directions,  $I({\bf q})$ at the particular  vector ${\bf q}$ can be  accurate enough to represent $I(q)$ at the magnitude $q$, without averaging over all directions of ${\bf q}$. A similar argument applies to bulk  liquids  or glasses, in which $I({\bf q})$ is also well self-averaged.

\subsection{Random Rotation of a Small Anisotropic Sample}

\label{sec:sq_rot}
The above method  of  using   a scattering vector ${\bf q}$ at {\em one} direction to represent  that magnitude $q$ does not work well for simulation samples, which are usually   small  and anisotropic (single crystal instead of polycrystal). 
To simulate experimental results, we can fix the direction of the incident ray but randomly rotate the small sample to many orientations.  This is done by applying a three-dimensional rotation matrix to  the original particle coordinates ${\bf r}_i$, whose rotation axis is uniformally distributed on a sphere and rotation angle is uniformally chosen from $[0, 2\pi]$. Then the signal $I({\bf q})$ in Eq.~(\ref{eq:Iq_exp}) can be approximated by accumulating intensities from all  those orientations $\Omega$
\begin{align}
\label{eq:Iq_rot}
I({\bf q}) =   \sum\limits_{\Omega}   \left[ \left |      \sum\limits_{i=1}^N    \cos({\bf q} \cdot {\bf r}_i (\Omega) ) \right|^2  +  \left |      \sum\limits_{i=1}^N     \sin({\bf q} \cdot {\bf r}_i(\Omega) ) \right|^2   \right].
\end{align}
Here, ${\bf r}_i(\Omega)$ represents the new coordinates of particle $i$ after  rotation to orientation $\Omega$.

 Note that the sum $ \sum\limits_{\Omega} $ is applied to the intensity rather than within the square like $\left |\sum\limits_{\Omega} \sum\limits_{i=1}^N   \cdots \right|^2$. The latter choice would imply a virtual system of many randomly orientated {\em overlapping} grains, each of size $N$.  The positions of these virtual grains generated by rotation do not reflect the absolute positions of grains in the real polycrystal.   According to the equivalency of  Eq.~(\ref{eq:Iq}) and Eq.~(\ref{eq:Iq_ij}),   Eq.~(\ref{eq:Iq_rot}) is an approximation to  Eq.~(\ref{eq:Iq_exp}) by only considering relative positions of particles {\em within}  each grain ${\bf r}_i(\Omega) - {\bf r}_j(\Omega) $. Therefore,  the difference between coordinates of particles $i$ and $j$ at two different orientations,  ${\bf r}_i(\Omega) - {\bf r}_j(\Omega') $, does not affect the result of Eq.~(\ref{eq:Iq_rot}), but will lead to different and wrong results, if the sum is taken as $\left |\sum\limits_{\Omega} \sum\limits_{i=1}^N   \cdots \right|^2$. 

For small and nearly isotropic liquids or glasses, one can replace random rotations of the sample by averaging over many thermally equilibrated configurations.  For anisotropic systems,  however,  rotations are needed to sample different directions. 

\subsection{Scattering Vector ${\bf q}$ on a  Lattice}

\label{sec:sq_lat}
An alternative and more convenient way to simulate experiments is to fix the sample  coordinates and choose ${\bf q}$ of a given $q$ from all directions. It is often suggested to select  ${\bf q}$  from a 3D orthorhombic lattice, ${\bf q} = \Delta q (n_x, n_y, n_z)$, with integers $n_x, n_y, n_z$ and   increment $\Delta q = \frac{2\pi}{L}$, where $L$ is the linear dimension of the cubic simulation box~\cite{allen1987}. The motivation here is that $L$ sets the maximum periodicity of the simulation sample that is still physically meaningful, and thus the resolution of $q$. The integers $n_x, n_y, n_z$  may be chosen to run from negative to positive  values to sample spherically symmetric ${\bf q}$'s, or to   start from zero to sample only ${\bf q}$'s on $1/8$ of the sphere. At the expense of   symmetry and averaging, the latter choice can reach  a higher magnitude $q$ with the same number of lattice points.

\begin{figure}
\centering
\includegraphics[width=0.45\textwidth]{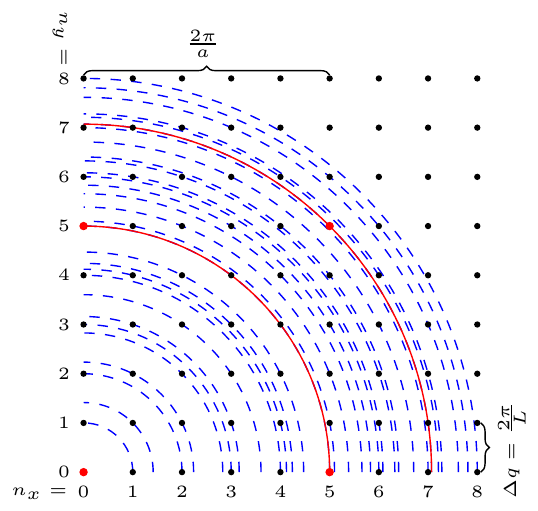}
\caption{Scattering vector ${\bf q}  = \Delta q (n_x, n_y)$ on a square lattice used for a 2D system of box size $L$.  ${\bf q} $ points at the same magnitude $q$ are connected by concentric quarter circles up to $q=8\Delta q$. If the system is a crystal of lattice constant $a = L/5$, then four points shown in red correspond to reciprocal lattice points.}
\label{fig:qgrid}
\end{figure}

There are multiple ${\bf q}$'s on this lattice that correspond to the same magnitude $q$, from which we can compute an average $S(q)$. The number of ${\bf q}$'s  for a given  magnitude $q$ tends to, but not necessarily, increase with $q$. For example, in a 2D system with ${\bf q}$'s on a square lattice, there are $1, 2, 1, 2, 2, 1, 2, \cdots$ ${\bf q}$ points on the lattice at magnitude $q/\Delta q = 0, 1, \sqrt{2}, 2, \sqrt{5}, \sqrt{8}, 9, \cdots $ respectively (Fig.~\ref{fig:qgrid}). When reporting the result of $S(q)$, one can  assign $q$'s into  bins of equal size or just use the original $q$ values visited by the lattice points. In both cases, the $S(q)$ should be the mean value averaged over all the ${\bf q}$'s at that $q$.

If the sample is crystalline and $L$ is an integer multiple of the crystallographic lattice constant $a$, then the ${\bf q}$ lattice   contains the reciprocal lattice points of the crystal (subject to a $2\pi$ factor difference)~\cite{allen1987}.  If $L = 5a$  in the above 2D example, then   ${\bf q}$ points $\Delta q(0,0)$, $\Delta q(5,0)$, $\Delta q(0,5)$, $\Delta q(5,5)$ correspond to reciprocal lattice points $(0,0)$, $(1/a, 0)$, $(0, 1/a)$, $(1/a, 1/a)$ respectively (Fig.~\ref{fig:qgrid}). These lattice points are where Bragg's law Eq.~(\ref{eq:bragg_q}) are obeyed. Therefore, according to the discussion in Section~\ref{sec:Fhkl}, if all atomic form factors are unity,  $I(q) = |F_{hkl}|^2=N^2$ and $S(q) = N$ at each of these reciprocal lattice points. The intensity scanning result $S(q)$ needs to be an average over all ${\bf q}$ points at that $q$, some of which are not  reciprocal lattice points and thus have $S(q)=0$.  For example, at $q = 5\Delta q$ of the 2D system, two points have $S(q) = N$ and two have $S(q)=0$. The average $S(q=5\Delta q)$ is thus $(N+N+0+0)/4 = N/2$ (Fig.~\ref{fig:qgrid}).

Using lattice points to approximate ${\bf q}$'s from all directions is problematic, when $L$ is small and thus increment $\Delta q$ is large such that only a few ${\bf q}$'s  are available at each $q$. The issue is more severe at small $q$ or towards corners of the cubic lattice at high $q$. The calculated signal $S(q)$   can   then  be quite noisy because ${\bf q}$ is not  averaged enough over all directions.

\subsection{Scattering Vector ${\bf q}$ on a  Sphere and  Debye's Scattering Equation }

\label{sec:sq_sph}
In order to obtain a smooth curve of $S(q)$ that better matches experimental results, we need to use enough  spherically distributed  ${\bf q}$'s. 
To guarantee uniform distribution of points on a sphere, we apply the Fibonacci grid approach to randomly chose $N_q$ scattering vectors ${\bf q}$'s  from a sphere of radius $q$~\cite{saff1997}. Increasing $N_q$ improves the effect of averaging. The  complexity  to compute $I(q)$ or $S(q)$ at each $q$ is then ${\mathcal O}(N_q N)$.

In the limit of $N_q \to \infty$,  using Eq.~(\ref{eq:Iq_ij}), we can integrate over all ${\bf q}$ directions  and then normalize it by the full solid angle of $4\pi$ to compute  the average $I(q)$ 
\begin{align}
\label{eq:Iq_debye}
\begin{array}{ll}
I(q)  &  =  \frac{1}{4\pi}  \int\limits_{|{\bf q}| = q} d{\bf q}   \sum\limits_{i=1}^N    \sum\limits_{j=1}^N   \hat{f}_i( q)  \hat{f}_j( q)  e^{ i {\bf q} \cdot  {\bf r}_{ij} }  \\
 &  =  \frac{1}{4\pi}  \int\limits_{0}^{2\pi} d\phi \int\limits_{0}^{\pi} \sin \theta d\theta    \sum\limits_{i=1}^N    \sum\limits_{j=1}^N   \hat{f}_i( q)  \hat{f}_j( q)  e^{ i  q r_{ij} \cos\theta}   \\
  &  =      \sum\limits_{i=1}^N    \sum\limits_{j=1}^N   \hat{f}_i( q)  \hat{f}_j( q)  \frac{\sin (q r_{ij})}{q r_{ij}}
\end{array}
\end{align}
This is known as   Debye's scattering equation~\cite{thomas2010,gelisio2016}, which can also be viewed as the discrete version of the Fourier transform of the radial distribution function $g(r)$ in Eq.~(\ref{eq:sq_gr}). The computational complexity of Debye's method is ${\mathcal O}(N^2)$ and it becomes more efficient than numerically sampling $N_q$ vector ${\bf q}$'s on a sphere when $N < N_q$.

\section{Photography and Intensity Scanning of Disordered or Partially Ordered Samples}

\label{sec:spec}
Although  intensity scanning, $I(q)$ or $S(q)$ as a function of $q$, generally gives more useful structural information about  isotropic samples, it is sometimes interesting to show the corresponding photography $I(X,Y)$.  In fact, intensity scanning can be obtained from  photography by moving along a specific radial direction on the $(X,Y)$ film, as in the early days of the powder method~\cite{cohen1935}.

To generate scattering photography of isotropic samples, we use the rotation or thermal averaging method of Section~\ref{sec:sq_rot}. Intensity scanning profiles are calculated using the three methods mentioned in Section~\ref{sec:sq_lat} and Section~\ref{sec:sq_sph}.

\subsection{Liquids and Glasses}

\begin{figure*}
\centering
\includegraphics[width=0.42\textwidth]{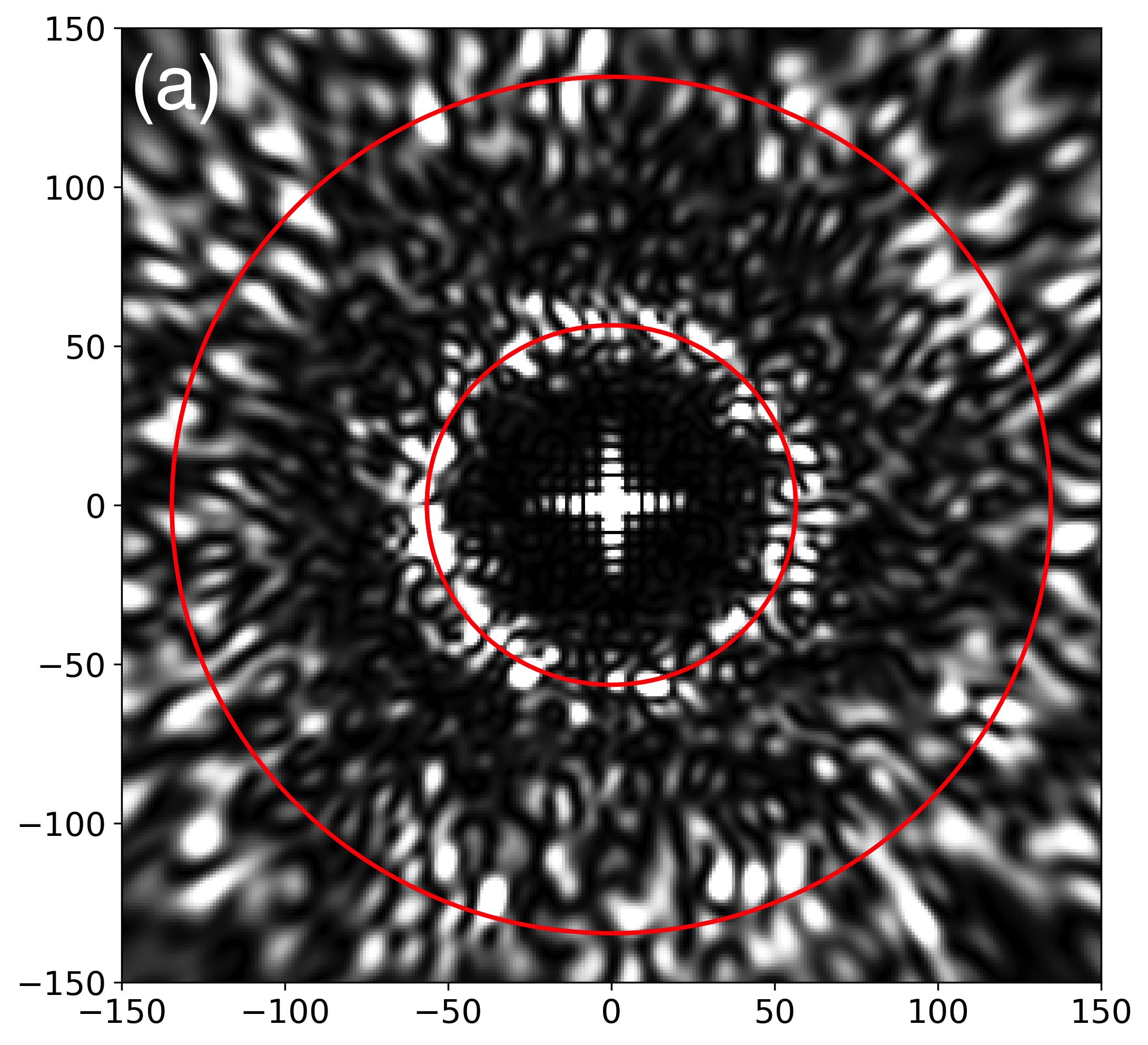}
\includegraphics[width=0.53\textwidth]{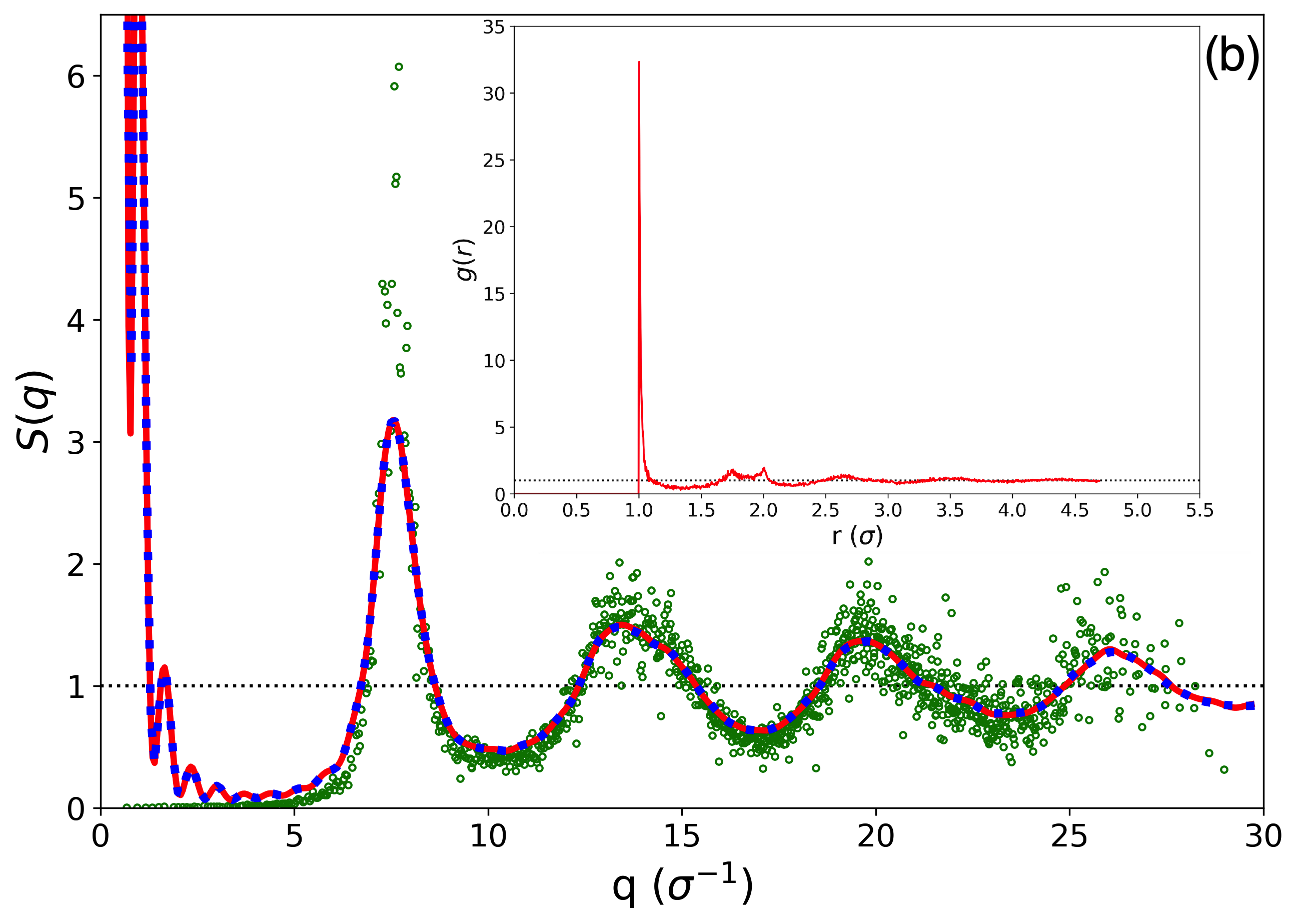}
\includegraphics[width=0.42\textwidth]{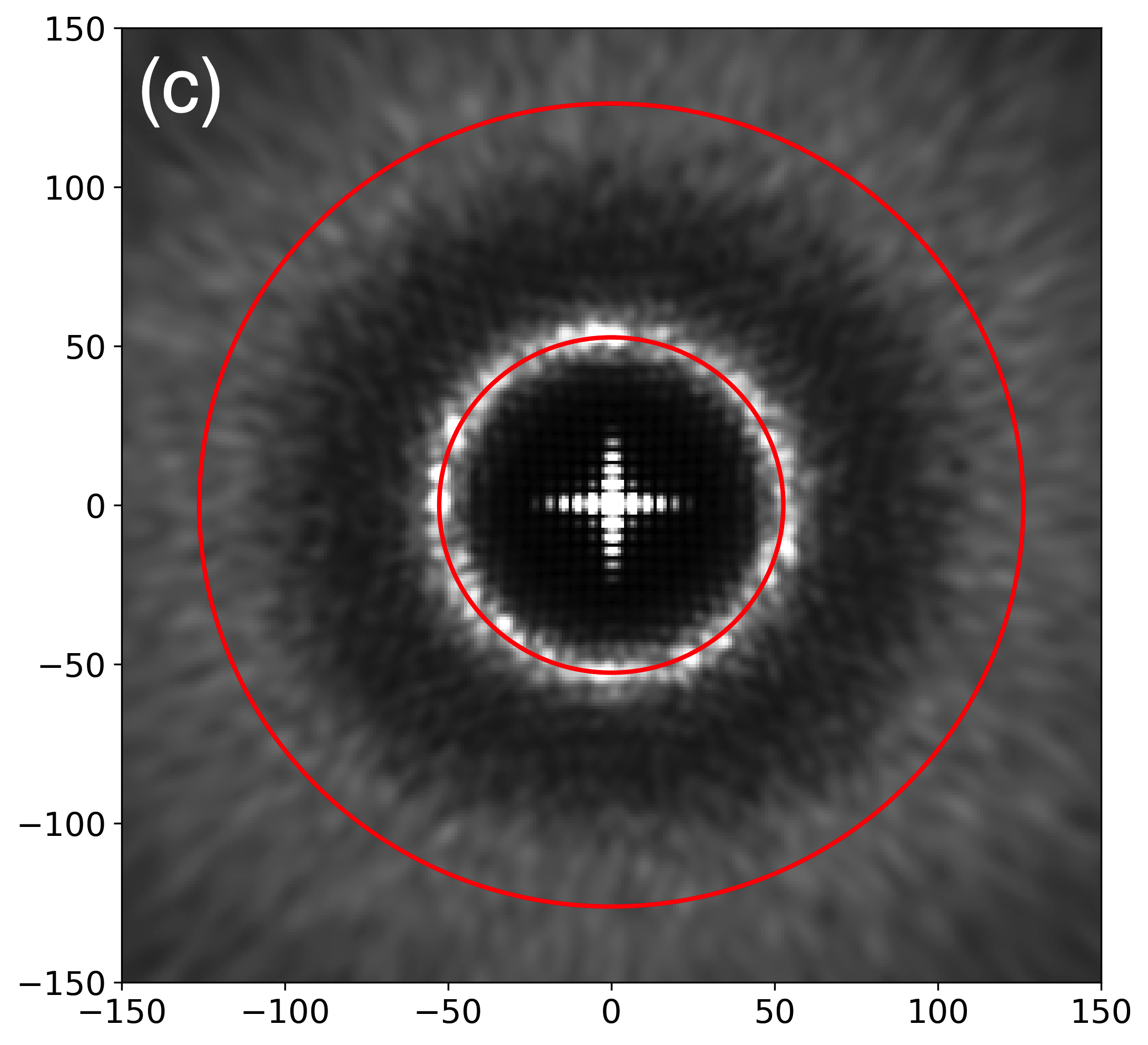}
\includegraphics[width=0.53\textwidth]{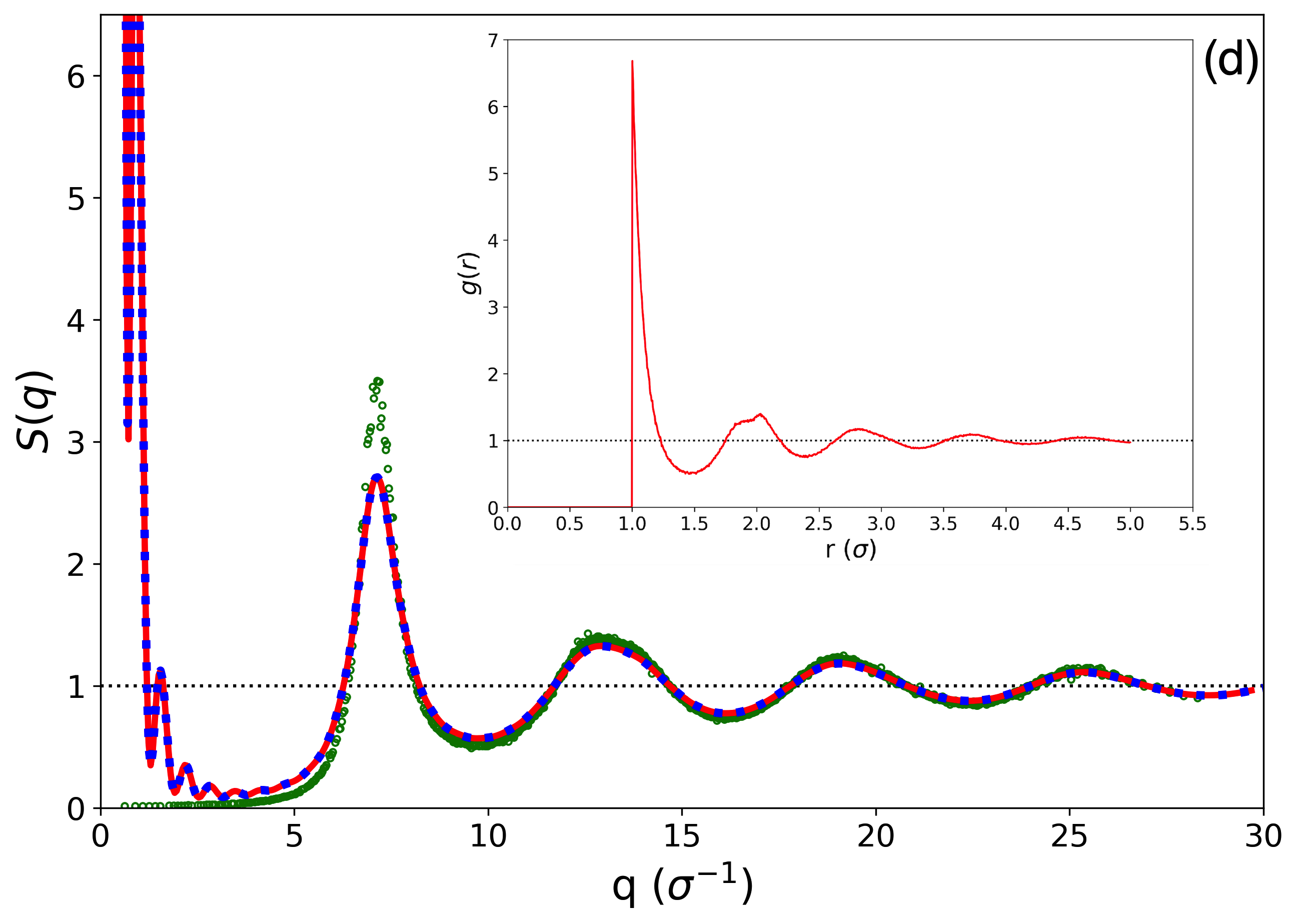}
\caption{Simulated transmission photography using fixed wavelength $\lambda=0.4\sigma$ (a,c) and structure factor $S(q)$ (b,d) of a   static glass sample (a-b) and a   thermally averaged liquid sample (c-d). Red solid rings in photography correspond to the first and second peaks in $S(q)$.  The cross pattern at the center of  each photograph is due to Fraunhofer diffraction from the  small  simulation box, effectively a cubic obstacle.  Three methods are used to compute $S(q)$:  with ${\bf q}$'s  on a cubic lattice  (green circles),  with ${\bf q}$'s on spheres (blue dotted line) and   Debye's scattering equation (red solid line).  Insets show the radial distribution function $g(r)$.   Both samples are $N = 1000$ hard spheres of diameter $\sigma$.  The glass sample has one configuration at  packing fraction $0.64$. The liquid sample has  1000  thermally equilibrated configurations  at packing fraction $\pi/6 = 0.5236$. }
\label{fig:sq_dis}
\end{figure*}

If a scattering photograph is taken for disordered samples like liquids or glasses using a fixed wavelength, a characteristic ring signal is expected at peak value $q^{*}\sim \frac{2\pi}{\sigma}$ that corresponds to the molecular size $\sigma$.  This ring  is  regular and clear,  when the sample, like most  experimental bulk samples,  is large enough such that a good average is taken within the system in the calculation of $I(\bf {q})$. However, in a small simulation system ($N=10^3$-$10^4$),  photography of one static disordered sample gives  spotty and noisy signals with certain traces of ring features (Fig.~\ref{fig:sq_dis}a). To enhance sharpness of the ring,  one can either increase the size $N$ of the sample or take  the ensemble average  of $I(\bf {q})$ over many configurations (Fig.~\ref{fig:sq_dis}c).

For homogeneous liquids and glasses, the static structure factor $S({\bf q}) = S(q)$ varies  only with the magnitude  $q$ of the scattering vector and  exhibits a major peak at $q^{*}\sim \frac{2\pi}{\sigma}$. Using ${\bf q}$'s on a sphere numerically or  Debye's equation  can generate well-averaged smooth $S(q)$ curves for liquids or glasses (Fig.~\ref{fig:sq_dis}b,d). If only one disordered configuration is analyzed,  the $S(q)$ curve is much noisier using  ${\bf q}$'s on a cubic lattice (Fig.~\ref{fig:sq_dis}b).

\subsection{Polycrystalline Samples -- Powder Method}

The powder method is often used to analyze polycrystals, in which a crystalline sample is ground into powder to produce many small randomly oriented crystalline grains. Then, at any scattering angle $2\theta$ where strong signal   is expected,  at least one of the grains has the correct orientation by chance to satisfy Bragg's law. The measured intensity scanning $S(q)$ can  be used to calculate interplanar spacings in the crystal and, with some limitations, even to determine  the crystal structure.

\begin{figure*}
\centering
\includegraphics[width=0.35\textwidth]{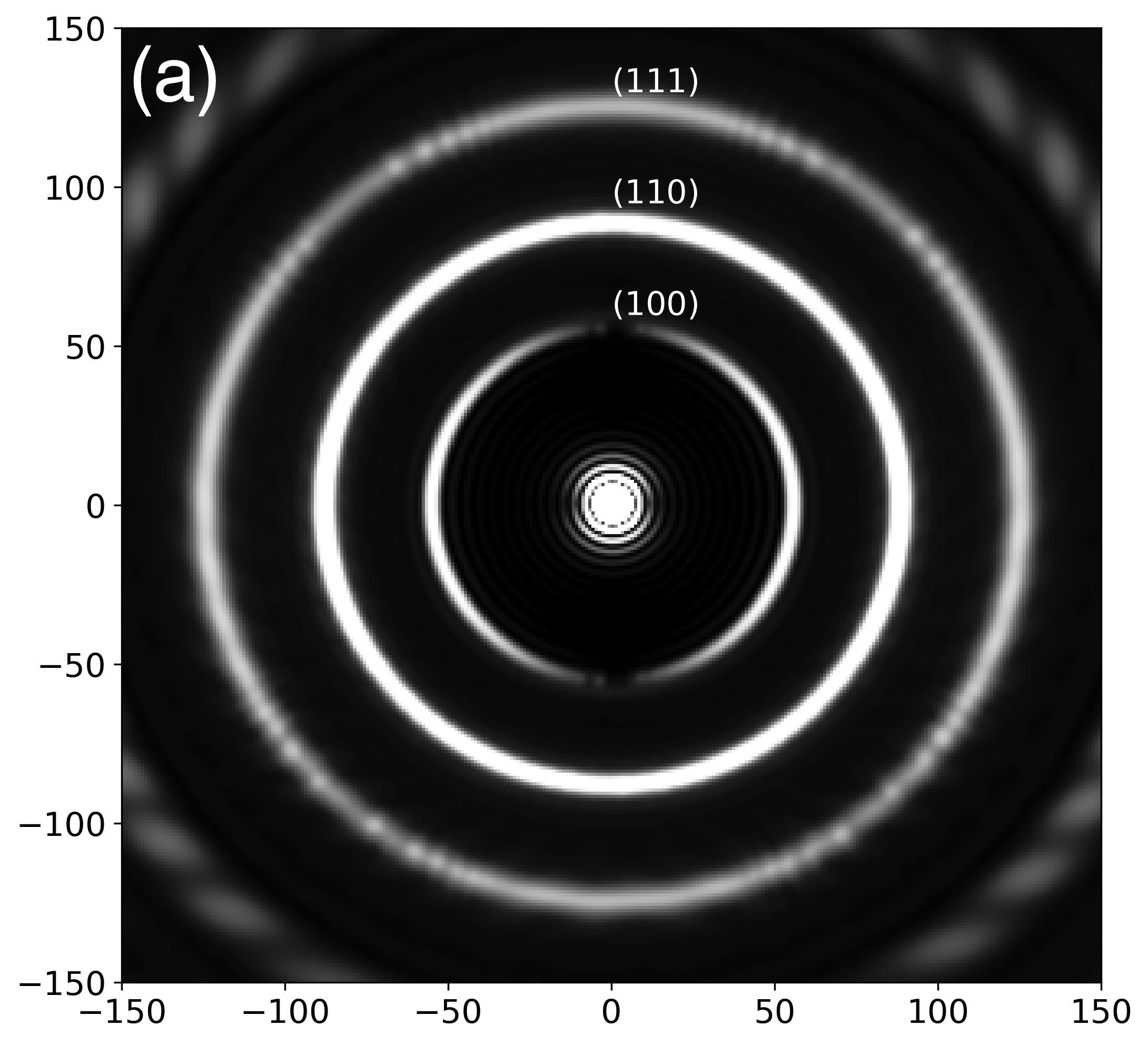}
\includegraphics[width=0.42\textwidth]{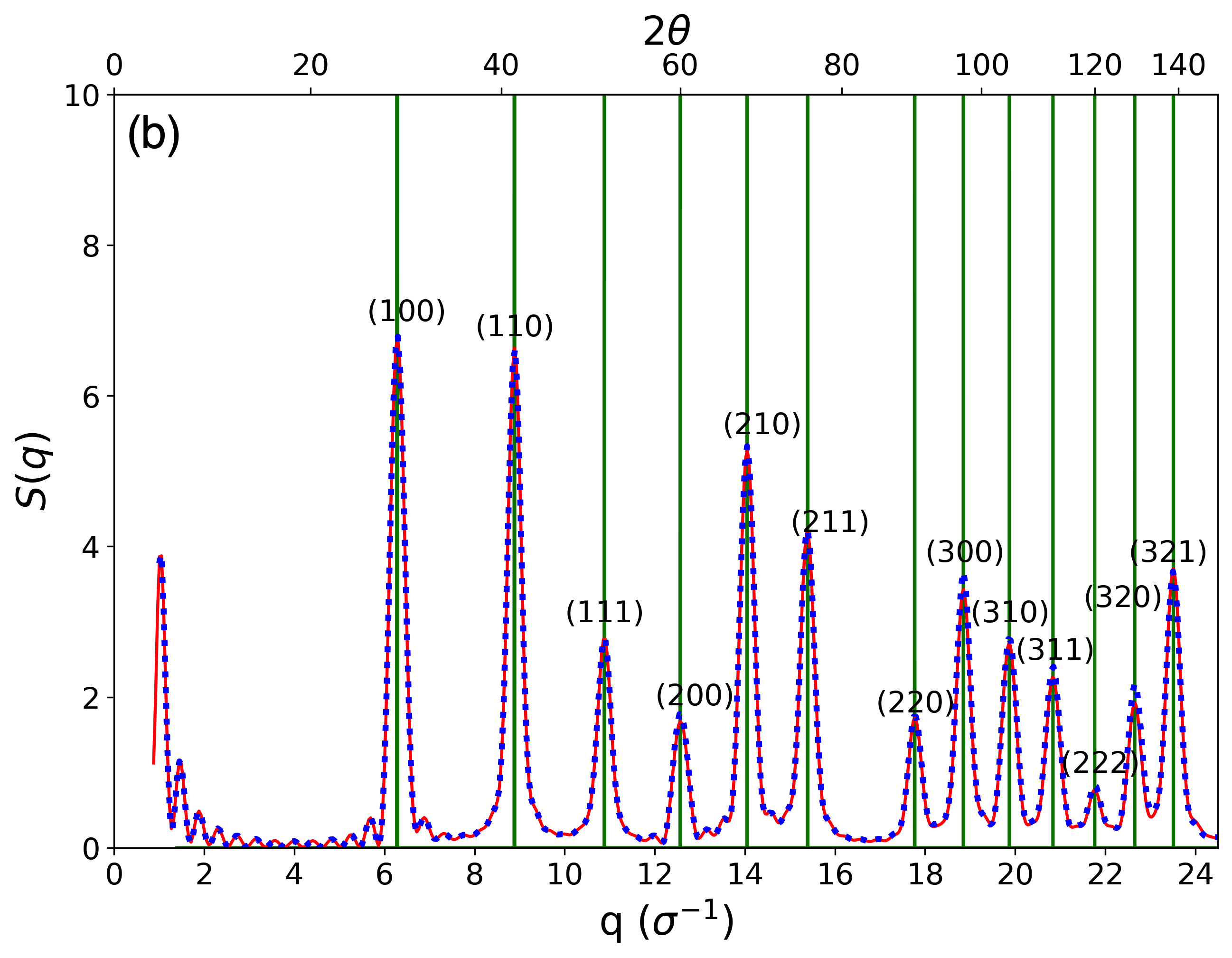}
\includegraphics[width=0.35\textwidth]{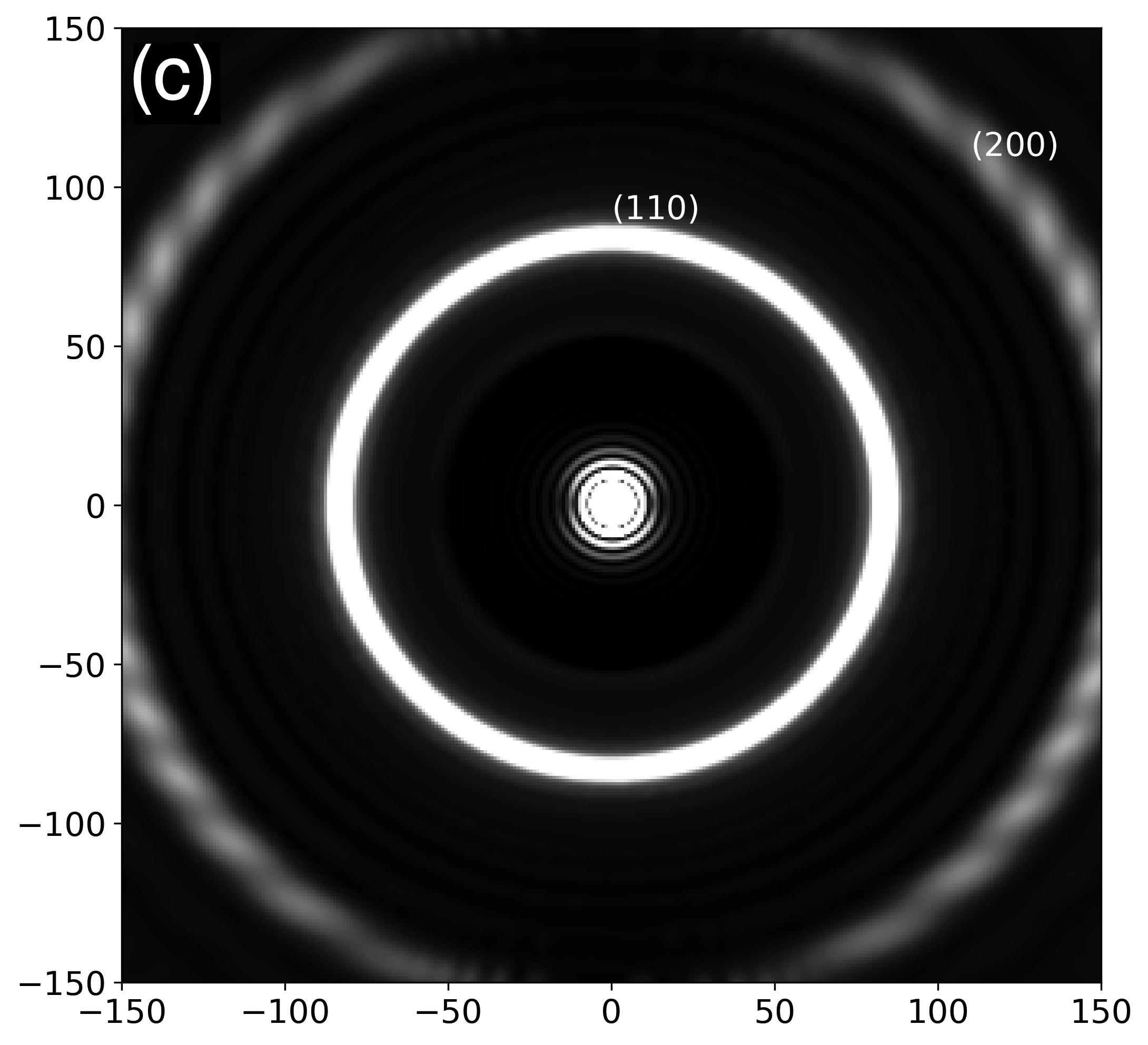}
\includegraphics[width=0.42\textwidth]{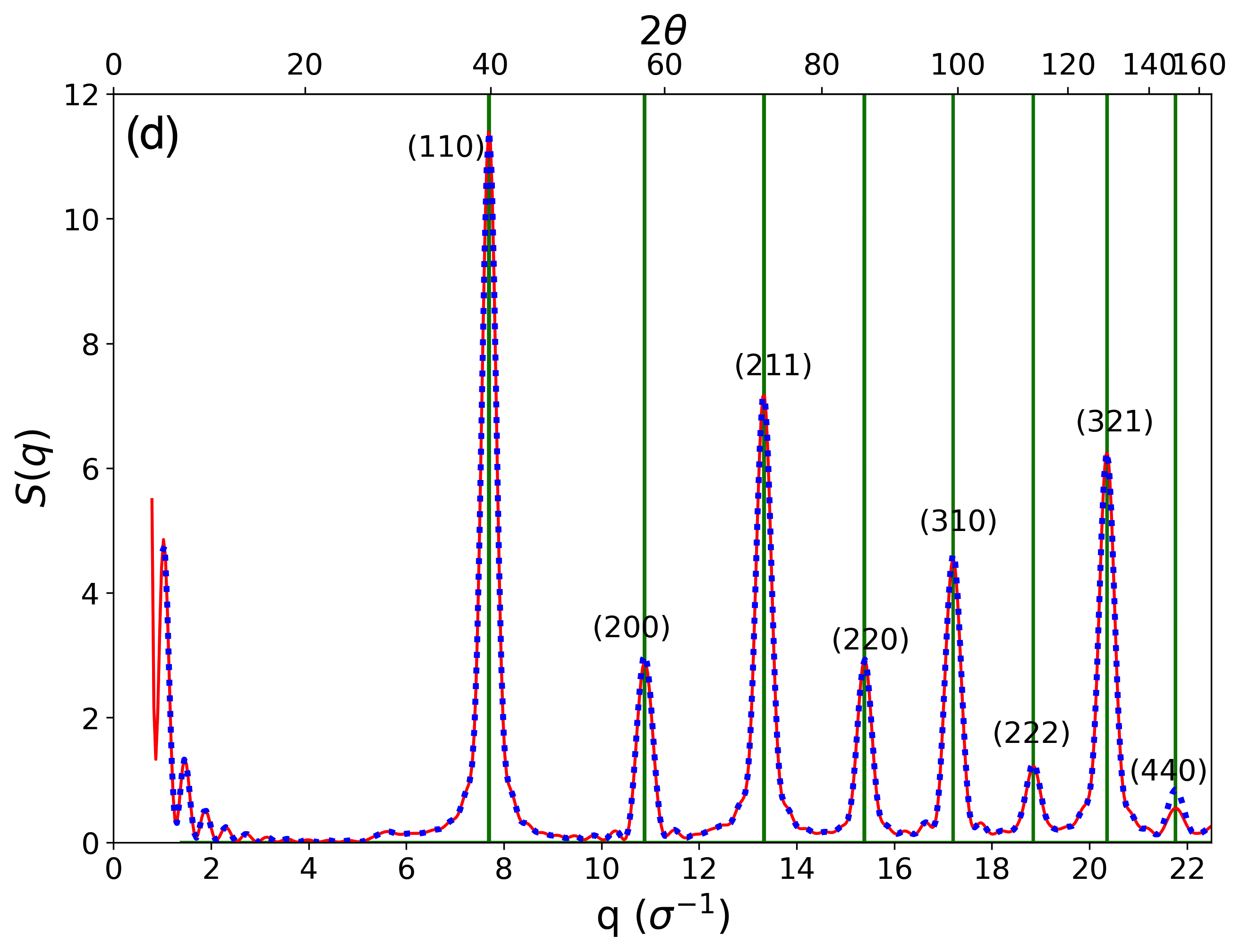}
\includegraphics[width=0.35\textwidth]{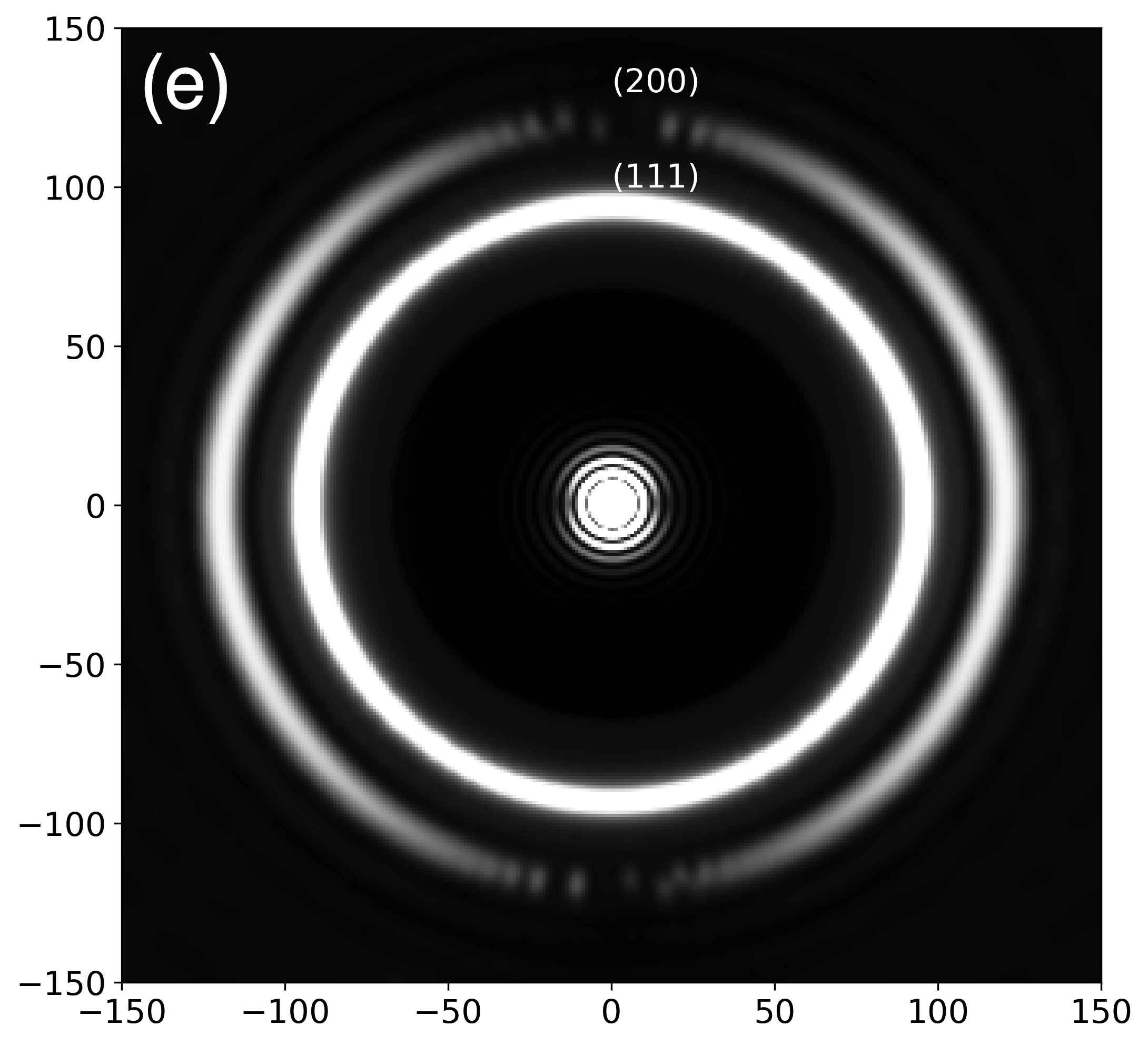}
\includegraphics[width=0.42\textwidth]{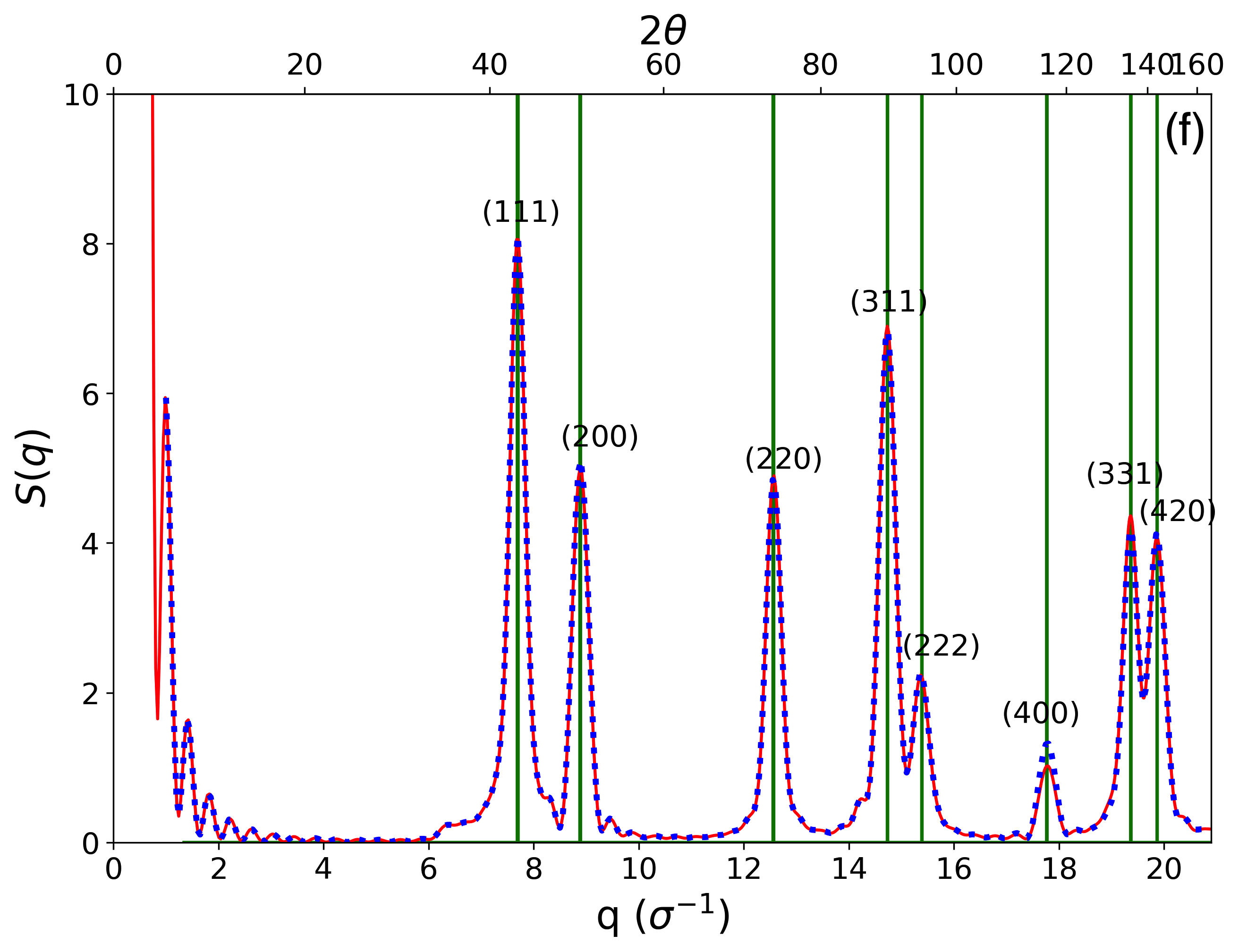}
\caption{Simulated powder method. Transmission photography  using fixed wavelength $\lambda=0.4\sigma$ (a,c,d) and structure factor $S(q)$ (b,d,f) of  polycrystalline SC (a-b), BCC (c-d) and FCC (e-f) samples.   Miller indices $(hkl)$ are labelled next to each signal peak.  The photograph is produced by randomly rotating a single crystalline sample in three dimensions and taking the average of $I({\bf q})$ over 5000 orientations. The small concentric circular pattern at the center of each photograph is due to Fraunhofer diffraction from the  small  simulation box, effectively  a circular obstacle, after being randomly rotated. Three methods are used to compute $S(q)$:  with ${\bf q}$'s  on a cubic lattice  (green vertical lines),  with ${\bf q}$'s on spheres (blue dotted line) and   Debye's scattering equation (red solid line).  }
\label{fig:powder}
\end{figure*}

It is difficult to produce a well randomized polycrystalline sample in simulation, given the limit of system size. Nevertheless, we can start from a small single crystal sample and use random rotation or {\bf q}'s from different directions to simulate scattering signals of a polycrystal. In particular, we use the same SC, BCC, FCC crystals used above to generate  photography and intensity scanning results of corresponding polycrystals (Fig.~\ref{fig:powder}).

The sharp concentric rings in the photograph $I(X,Y)$ and the narrow peaks in $S(q)$ correspond to scattering from different crystallographic planes $(hkl)$ of the three crystals (SC, BCC, and FCC). $S(q)$ peaks computed from spherically distributed ${\bf q}$'s are lower and broader than those from cubic lattice  ${\bf q}$'s. The peak height  using cubic lattice  ${\bf q}$'s often scales with system size $N$. For example, the SC crystal has $L = 15\sigma$  and $N = L^3 = 3375$ particles.  Given $\Delta q = \frac{2\pi}{15\sigma}$,  the $S(q)$ peak from $(100)$ planes is expected to occur at six ${\bf q}$ points, $\Delta q (15,0,0)$, $\Delta q (0,15, 0)$, $\Delta q (0,0,15)$, $\Delta q (-15,0,0)$, $\Delta q (0, -15,0)$, and $\Delta q (0,0,-15)$, each has a value $S(q)=3375$. However, there are other ${\bf q}$ points with magnitude $q=15\Delta q$, which correspond to integer solutions to $n_x^2 + n_y^2 + n_z^2 = 15^2$.  In total, at $q=15$, there are 6 $(15,0,0)$-like (considering its permutation and $\pm$),  24  $(12, 9, 0)$-like, 
24 $(10, 10, 5)$-like,   48 $(11,10,2)$-like, and  48 $(14,5,2)$-like ${\bf q}$  points. Out of these 150 points, only 6 have $S(q)=N$ while others have $S(q)=0$. So the peak height $S(q=15\Delta q) = 3375\times 6/150 = 135$.

\subsection{Mesophases -- Small-Angle Method}

Mesophases are states of matter intermediate between liquids and solids found in block copolymers~\cite{sakurai1991}, liquid crystals~\cite{mitchell1983},  structural DNAs~\cite{tian2020}, etc., which present mesoscopic ordering of length scales larger than molecular size $\sigma$.  To detect these long-wavelength structures,  small-angle X-ray scattering (SAXS)~\cite{chu2001} or small-angle  neutron scattering (SANS) ~\cite{richards1983}  methods  are needed because scattering signals are expected at  small $q$ (before the first major diffraction peak $\sim\frac{2\pi}{\sigma}$) thus small $\theta$ as seen from Eq.~(\ref{eq:qtheta}).  A logarithmic scale axis is often set for $S(q)$ in the structure factor plot because at $q \to 0$ the signal scales with system size $N$~\cite{schneidman2010}.

Experimental mesophases are often synthesized as polycrystals or many small crystalline domains randomly embedded in an amorphous matrix, for which intensity scanning $S(q)$ at small $q$ is used to reveal the ordering. We illustrate the concepts of the mesophase structure factor using a lamellar, a cylindrical and a BCC spherical configuration of domains,  cut from a disordered glass sample of hard spheres of diameter $\sigma$.  These structures are thus amorphous within each domain, but the domains form a 1D, 2D or 3D superlattice for the lamellar, cylindrical or spherical configuration, respectively.

\begin{figure}
\centering
\includegraphics[width=0.49\textwidth]{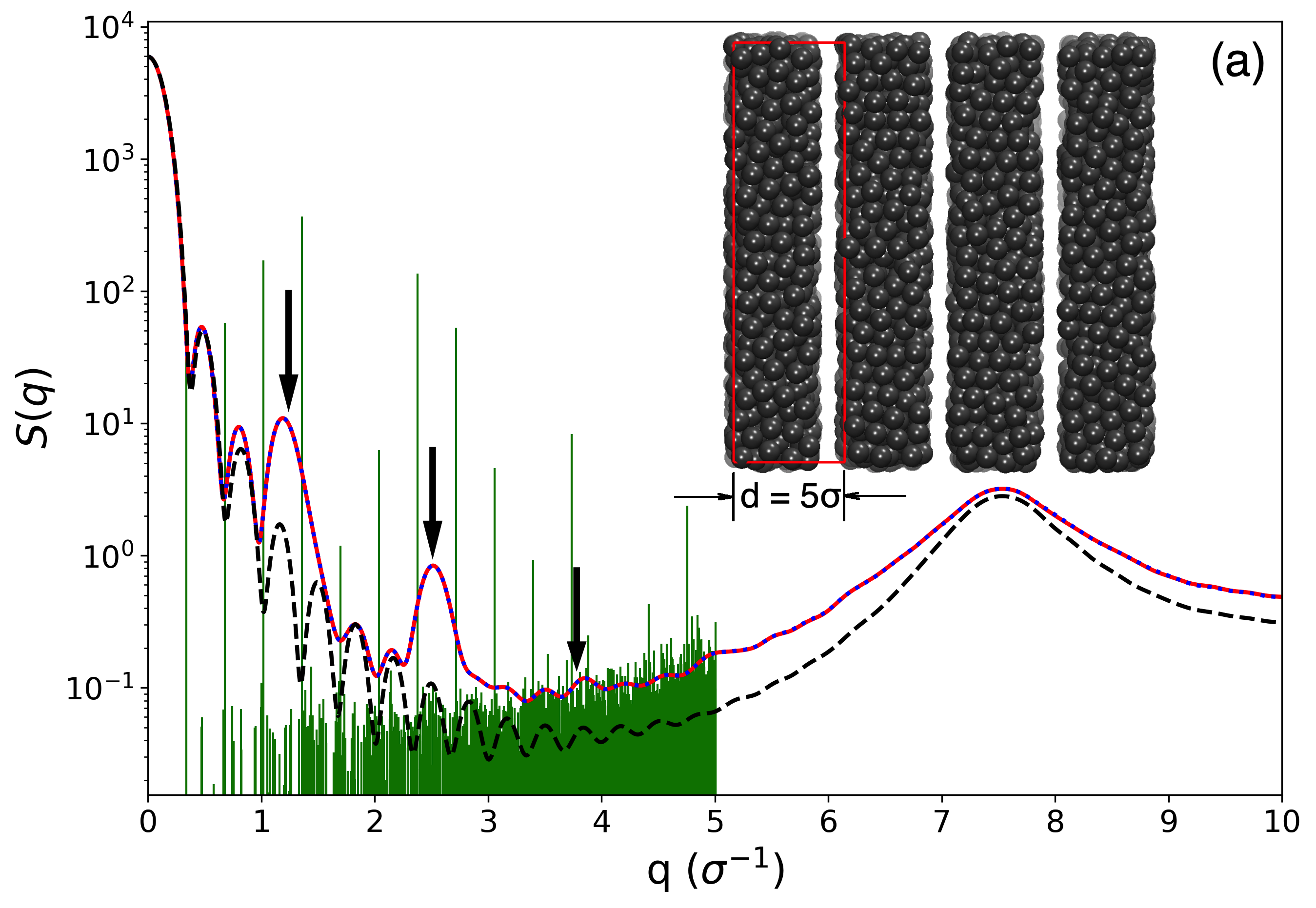}
\includegraphics[width=0.49\textwidth]{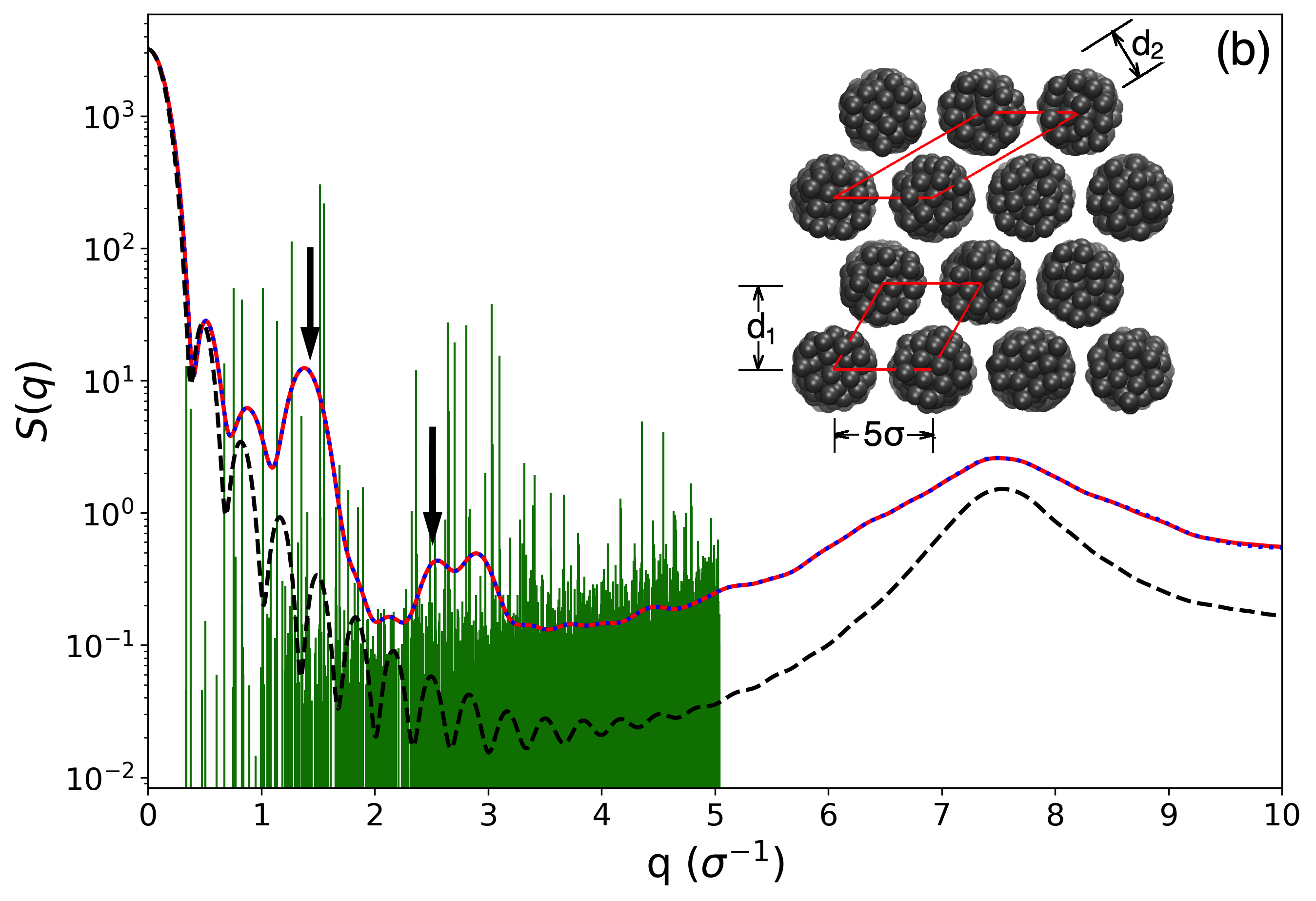}
\includegraphics[width=0.49\textwidth]{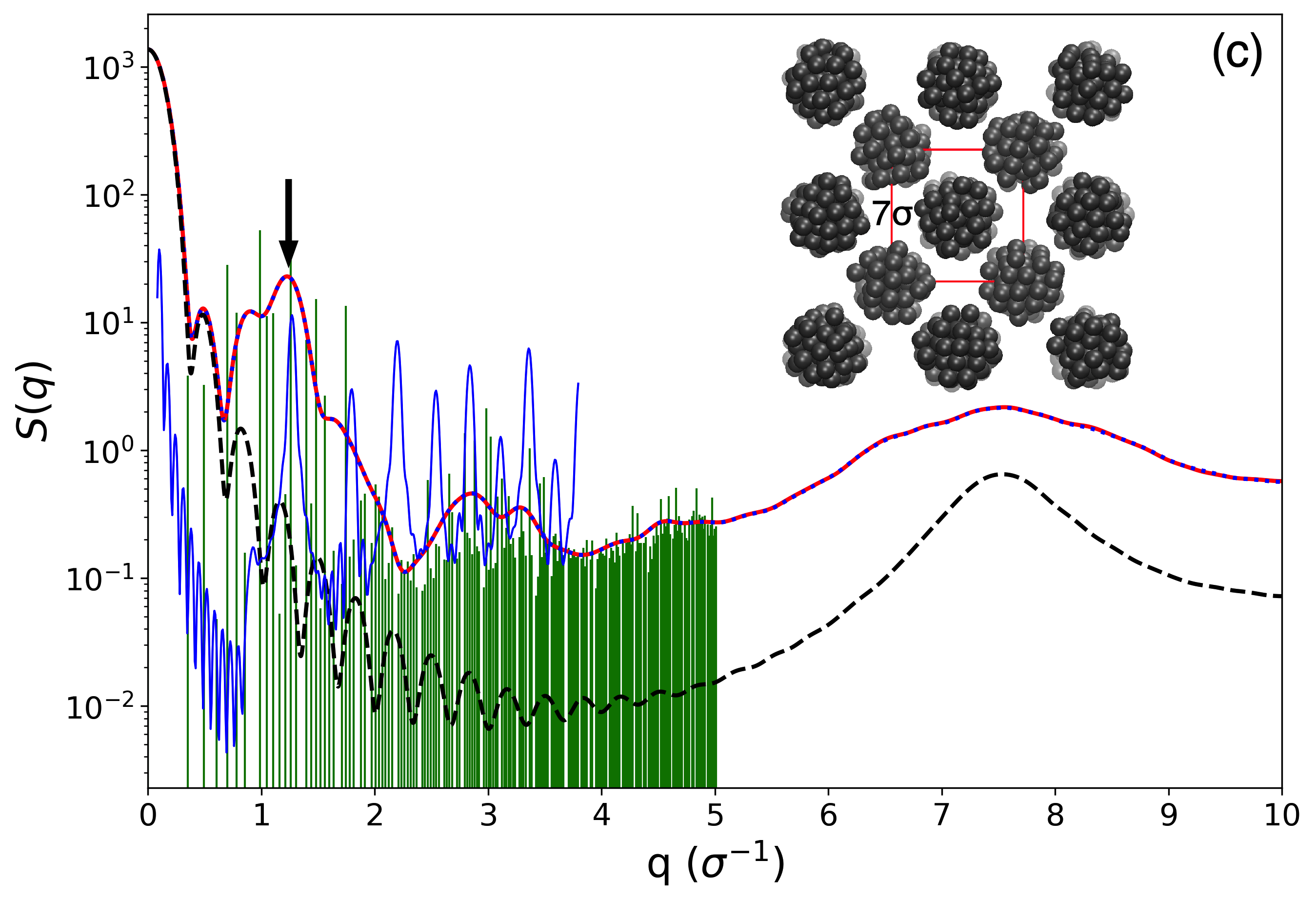}
\caption{Simulated small-angle structure factor $S(q)$ for (a) lamellar, (b) cylindrical and (c) BCC spherical mesophases.  Three methods are used to compute $S(q)$:  with ${\bf q}$'s  on a cubic lattice  (green vertical lines),  with ${\bf q}$'s on spheres (blue dotted line) and   Debye's scattering equation (red solid line).   Debye's result of  $S(q)$  of a homogeneous glass, after vertically rescaled to align at $q\to 0$,   is shown for comparison (black dashed line). Black downward arrows mark signature peaks for each structure. The broad peak at $7.5\sigma^{-1}$ corresponds to particle size $\sigma$. Insets show top/side views of the configurations under consideration.  Blue solid line in (c) is Debye's result of  $S(q)$ for a BCC sphere mesophase with one particle per domain, obtained by rescaling the curve of a BCC crystal.  }
\label{fig:meso}
\end{figure}

In the lamellar phase,  each period is of length $d = 5\sigma$ consisting of a layer with thickness $3.5\sigma$ and a gap with thickness $1.5\sigma$. We find three peaks of $S(q)$ at one time, two times and three times  $\frac{2\pi}{d}\approx 1.257 \sigma^{-1}$, corresponding to the first, second and third order of Bragg diffraction of the superlattice (Fig.~\ref{fig:meso}a). The peak height drops as $q$ increases,  and when layer thickness equals the gap thickness,  peaks at  even multiples of $\frac{2\pi}{d}$ disappear.

The cylindrical phase with a disk radius $1.8\sigma$ resides on a two-dimensional triangular superlattice with  lattice constant $5\sigma$.  By assigning unit cells in two different ways with interplanar spacing $d_1 = \frac{5\sqrt{3}}{2} \sigma$ and $d_2 = 2.5\sigma$, we can identify two peaks at  $q_1 = \frac{2\pi}{d_1}  \approx 1.451 \sigma^{-1}$ and $q_2 = \frac{2\pi}{d_2} \approx 2.513 \sigma^{-1} $ (Fig.~\ref{fig:meso}b). 
The second-order peak  around  $2 q_1  \approx 2.9 \sigma^{-1}$ is also visible (not marked).

The spherical phase has spheres of radius $2\sigma$ that pack on a BCC superlattice with a lattice constant $7\sigma$.  If each sphere domain has only one particle, the structure factor would be the same as a normal BCC crystal apart from a change of  unit  for $q$. We can obtain $S(q)$ of this one-particle  spherical phase by rescaling the $q$ axis  of $S(q)$ of the BCC crystal, which has a lattice constant  $a = \frac{2}{\sqrt{3}}\sigma$,  by a factor of  $\frac{7\sqrt{3}}{2} \approx 6.062$. This moves the $(110)$ peak from $7.7\sigma^{-1}$ to $1.27\sigma^{-1}$ (Fig.~\ref{fig:meso}c). This helps us to identify that only the peak from the $(110)$ planes of the BCC superlattice is sharply distinguishable from the background signals.

\section{2D Structure Factor}

\label{sec:2Dsq}
For 2D samples or 2D projection of 3D samples, it is sometimes useful to express $S({\bf q})$ as a 2D function of $(q_x,q_y)$~\cite{tutsch2014} or scattering angles $(\theta_x, \theta_y)$~\cite{lee2005}. The 2D structure factor $S(q_x, q_y)$ is related to the photography $I(X,Y)$ by converting coordinates $(X,Y)$ on the film into components $(q_x,q_y)$ of the scattering vector using   Eq.~(\ref{eq:qxqy}). For 3D structures, the component $q_z$ can be expressed as a function of $q_x$ and $q_y$, for example, in the case of  the transmission method (Eq.~(\ref{eq:q_trans})),
\begin{align}
\label{eq:qz}
 \begin{aligned}
q_z & =   -  \frac{2\pi}{\lambda} (1 -  D/L) \\
& = -  \frac{2\pi}{\lambda} \left (1 - \sqrt{1 -  \frac{\lambda^2}{(2\pi)^2}q_x^2 - \frac{\lambda^2}{(2\pi)^2} q_y^2} \right )  \\
& =  -  \frac{2\pi}{\lambda}  + \sqrt{ \left(  \frac{2\pi}{\lambda} \right)^2 - q_x^2 - q_y^2}.
\end{aligned} 
\end{align}
Note that knowing $(q_x, q_y)$ does not uniquely determine $q_z$. The constant $\frac{2\pi}{\lambda}$ still needs to be specified.
For small-angle scattering with $q_x,q_y \to 0$, an approximation to set $q_z = 0$ is  valid if there is no long-range periodicity along the $z$ direction.

\begin{figure}
\centering
\includegraphics[width=0.45\textwidth]{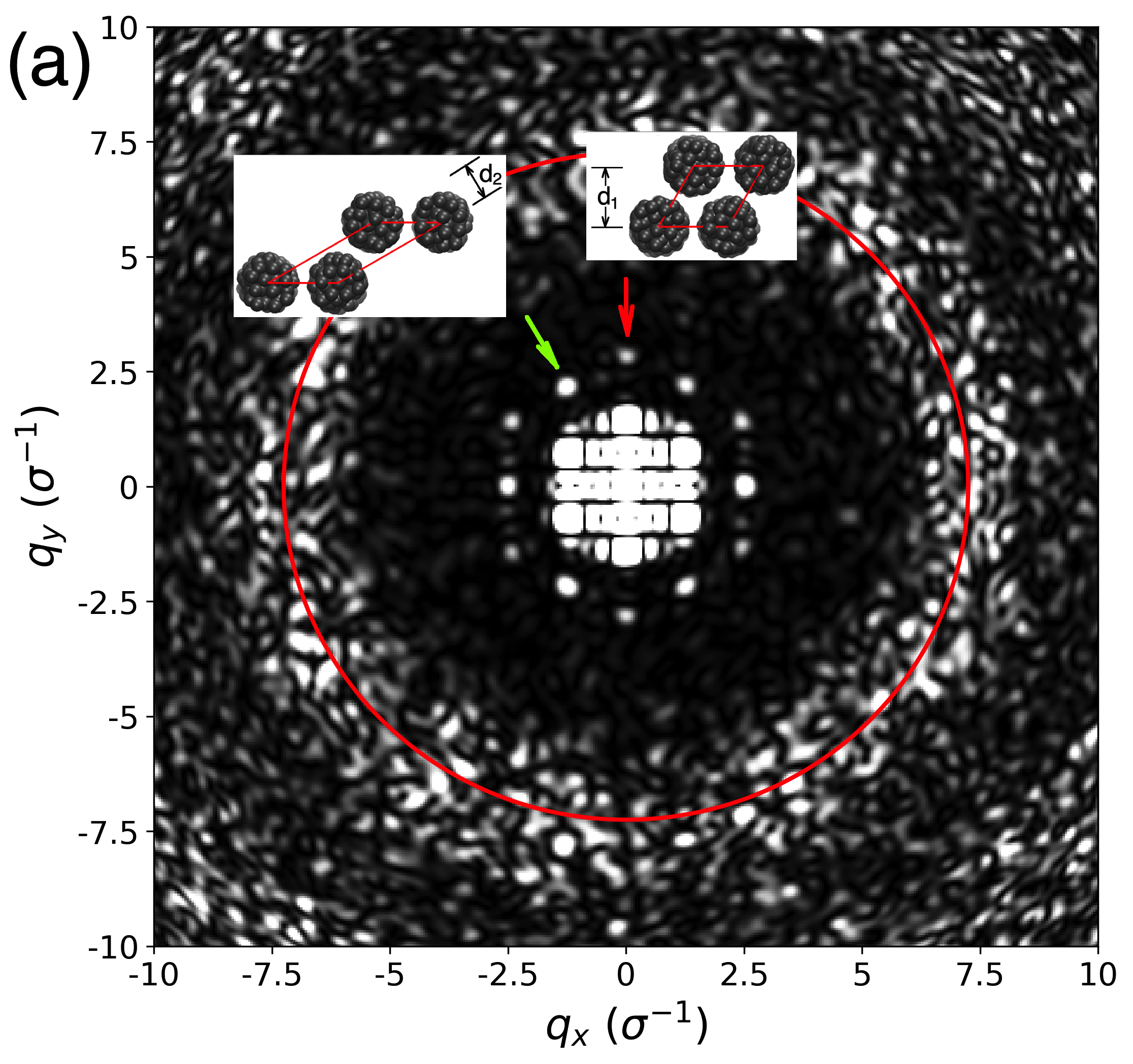}
\includegraphics[width=0.45\textwidth]{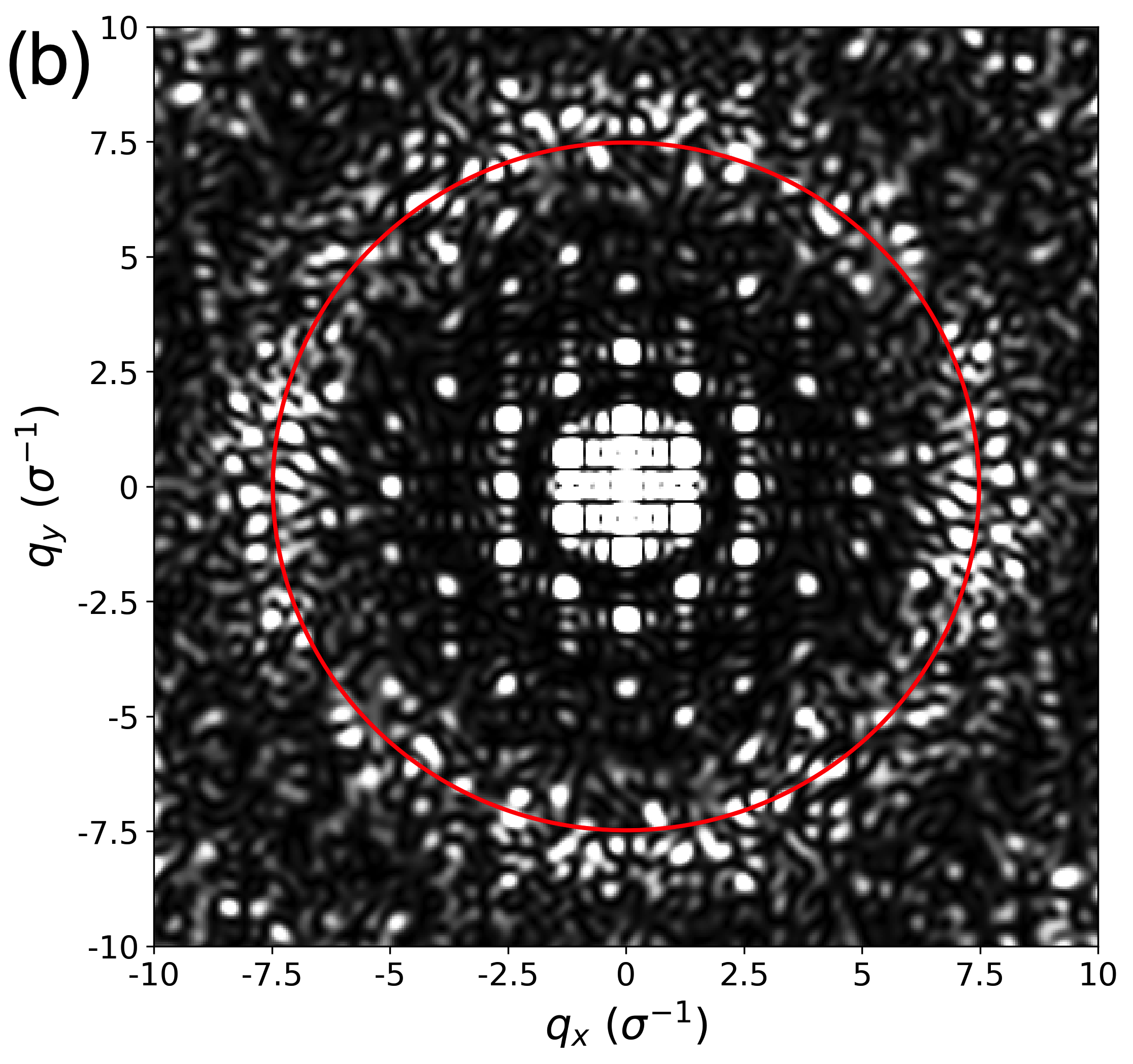}
\caption{2D structure factor $S(q_x, q_y)$ of the cylindrical mesophase using $\lambda = 0.4\sigma$ with (a) $q_z = -  \frac{2\pi}{\lambda}  + \sqrt{ \left(  \frac{2\pi}{\lambda} \right)^2 - q_x^2 - q_y^2}$ and (b) $q_z = 0$. The red solid ring marks the broad peak corresponding to particle diameter $\sigma$ at $\sim \frac{2\pi}{\sigma}$.  The red arrow points to the second-order peak from the unit cell with spacing $d_1$. The green arrow points to the first-order peak from the unit cell with spacing $d_2$. }
\label{fig:Sxy}
\end{figure}

We compute $S(q_x, q_y)$ for the cylindrical mesophase in Fig.~\ref{fig:meso}b, whose cylinder axis is aligned with the incident ray in the $z$  direction. We first use $q_z$ calculated from Eq.~(\ref{eq:qz}) with $\lambda = 0.4\sigma$. Besides the isotropic circular signal corresponding to particle size $\sigma$,  a characteristic hexagonal pattern with six-fold symmetry is observed at small $q$, which results from the cylinders packed on a 2D triangular lattice. We can identify two sets of spots on the vertices of hexagons -- one corresponds to the unit cell with spacing $d_1$ and the other corresponds to the unit cell with spacing $d_2$ (Fig.~\ref{fig:Sxy}a). The second-order peak related to $d_1$ and first-order peak related to $d_2$ form a hexagon together, while the first-order peak related to $d_1$ is   mixed with the Fraunhofer diffraction pattern at smaller $q$.

If we set $q_z = 0$, the $S(q_x,q_y)$ pattern is approximately the same at small $q$, with a certain degree of enhancement  (Fig.~\ref{fig:Sxy}b). Some   higher-order peaks become visible at larger $q$.

\section{Conclusion}
\label{sec:con}
In this article, we give a comprehensive and coherent review of core concepts about scattering methods used to determine the structures of ordered and disordered samples.  Scattering photography and intensity scanning of typical examples are calculated that can be used as benchmarks. Sample CPU codes are provided on GitHub at \url{https://github.com/statisticalmechanics/scatter} to illustrate the mathematics and algorithms. Accelerating  GPU codes that can reduce hours of computation to seconds are also provided for efficient simulation of scattering signals. It should be noted that for simplicity, the intensity calculation discussed in this paper has omitted serveral important wavelength and/or angle-dependent factors due to, for example, absorption, extinction, multiple scattering, polarization, and the Lorentz factor, as well as the issue of normalization of the measured intensity to an absolute scale. We have also omitted the effect of temperature which generally adds a Gaussian co-factor to each atomic scattering factor.

\begin{acknowledgments}
This work benefits from the  Duke Kunshan  startup funding and resources made available at the Duke Compute Cluster (DCC). We thank  Corey O'Hern, Alex Grigas, Robert Hoy and Joseph Dietz for testing some of our codes. We also thank Patrick Charbonneau for helpful discussions.
\end{acknowledgments}

\begin{appendix}
	
\section{Fourier Transform: Continuous and Discrete}

\label{sec:ft}
The Fourier transform $\hat{F}_{\bf k}$ of  a function $F({\bf r})$ defined  continuously in three-dimensional real space of infinite volume  is
\begin{align}
\label{eq:fourier}
\hat{F}_{\bf k} = \int  d{\bf r}  F({\bf r})  e^{ i {\bf k} \cdot {\bf r}},
\end{align}
where ${\bf k}$ is a wavevector used to extract the spatial periodicity of $F({\bf r})$~\cite{lighthill1958}. For instance,  if $F({\bf r})$ has a periodic pattern of wavelength $\lambda$ along the $x$ axis, i.e. $F(x,y,z) = F(x + \lambda, y, z)$, then the value of $\hat{F}_{\bf k}$ is large for the ${\bf k}$  of magnitude $|{\bf k} | = 2\pi / \lambda$ pointing in the $x$ direction, i.e.  ${\bf k} = (2\pi / \lambda, 0, 0)$.  Physically, if  $e^{ i {\bf k} \cdot {\bf r}} $ is viewed as a plane wave traveling in the  ${\bf k}$ direction, then $\hat{F}_{\bf k}$ would exhibit a peak value, when $F({\bf r})$ has wavelike properties coherent with  $e^{ i {\bf k} \cdot {\bf r}} $ such that they add    constructively in the integral. In this sense, the Fourier transform Eq.~(\ref{eq:fourier}) quantifies the existence and the extent of periodicity  corresponding to ${\bf k}$ in $F({\bf r})$.

In general, even if $F({\bf r})$ is a real function,  $\hat{F}_{\bf k} $ can be  complex. However, if  $F({\bf r})$  is real ($F^*({\bf r}) = F({\bf r}) $) and even ($F(-{\bf r}) = F({\bf r})$, i.e. with a symmetry center),   its Fourier transform $\hat{F}_{\bf k} $ is also  real and even, because the conjugate of $\hat{F}_{\bf k} $ is
\begin{align*}
\begin{aligned}
\hat{F}^*_{\bf k}  &= \int  d{\bf r}  F^*({\bf r})  e^{- i {\bf k} \cdot {\bf r}}  = \int  d{\bf r}  F({\bf r})  e^{- i {\bf k} \cdot {\bf r}}  = \hat{F}_{-\bf k}  \\
&= \int_{-\infty}^{\infty}  d{\bf r}  F({-\bf r})  e^{- i {\bf k} \cdot {\bf r}}     \stackrel{{\bf r}' = -{\bf r}}{=}  - \int_{\infty}^{-\infty}   d{\bf r}'  F({\bf r}')  e^{ i {\bf k} \cdot {\bf r}'}  \\
&=    \int_{-\infty}^{\infty}   d{\bf r}'  F({\bf r}')  e^{ i {\bf k} \cdot {\bf r}'} 
 = \hat{F}_{\bf k}.
\end{aligned}
\end{align*}
Here  the integration limits for the vector variable ${\bf r}$, formally denoted as $\pm \infty$, are to be understood as for each of its components.

The inverse Fourier transform of $\hat{F}_{\bf k}$ is an integral in the wavevector space which  gives the original real-space function
\begin{align}
\label{eq:fourinv}
F({\bf r})  = \frac{1}{(2\pi)^3} \int  d {\bf k}  \hat{F}_{\bf k} e^{- i {\bf k} \cdot {\bf r}}.
\end{align}
This expands $F({\bf r})$ in terms of an infinite number of  periodic basis functions $ e^{- i {\bf k} \cdot {\bf r}}$ characterized by different ${\bf k}$'s. The coefficient or contribution of each ${\bf k}$ is just the Fourier transform $\hat{F}_{\bf k}$. In principle, the collection of all $ \hat{F}_{\bf k}$'s contains  the entire information about the original function $F({\bf r})$ such that knowing $\hat{F}_{\bf k}$'s  allows us to reconstruct $F({\bf r})$.

In physical systems, $F({\bf r})$  is often defined within a finite volume $V$ and  the Fourier transform should be integrated over  the region $V$
\begin{align}
\label{eq:fourV}
\hat{F}_{\bf k} = \int\limits_V  d{\bf r}  F({\bf r})  e^{ i {\bf k} \cdot {\bf r}}.
\end{align}
If such a finite system is of a cubic shape with  a linear dimension $L$, i.e. $V= L^3$,  then any periodicity or wavelength $\lambda > L$ is unphysical. This imposes a lower bound, $2\pi/L$, on the smallest wavevector to be considered. The inverse Fourier transform Eq.~(\ref{eq:fourinv}) thus should not   vary ${\bf k}$  continuously as in an integral, but only take discrete  values of ${\bf k}$  with increments $(\Delta k_x, \Delta k_y, \Delta k_z )  = ( \frac{2\pi}{L}, \frac{2\pi}{L}, \frac{2\pi}{L})$. The integral then becomes~\cite{chaikin1995}
\begin{align}
\label{eq:fourVinv}
F({\bf r}) =  \frac{1}{(2\pi)^3} \sum_{{\bf k}} \hat{F}_{\bf k} e^{- i {\bf k} \cdot {\bf r}} \left(\frac{2\pi}{L} \right)^3 = \frac{1}{V} \sum_{{\bf k}} \hat{F}_{\bf k} e^{- i {\bf k} \cdot {\bf r}}.
\end{align}

Mathematically, for the Fourier transform Eq.~(\ref{eq:fourier}) to exist, the function $F({\bf r})$ needs to be absolutely  integrable. If  $F({\bf r})$ equals to some nonzero constants, or without loss of generality, $F({\bf r})  = 1$,  in order to reconcile the singularity, the result of the  Fourier transform is formally written as $(2\pi)^3\delta_D({\bf k})= \int  d{\bf r}   e^{ i {\bf k} \cdot {\bf r}}$, or equivalently,
\begin{align}
\label{eq:dirac}
\delta_D({\bf k})=  \frac{1}{(2\pi)^3} \int  d{\bf r}   e^{ i {\bf k} \cdot {\bf r}},
\end{align}
where $\delta_D({\bf x})$ is the (three-dimensional) singular Dirac delta function ($\delta_D(0) \to \infty$). Usually,  the Dirac delta function with the property that $\int  d {\bf x} \delta_D({\bf x}) f({\bf x}) = f(0)$ is introduced as the limiting case of a normalized Gaussian function with  its standard deviation approaching zero. According to the above notation, the inverse Fourier transform of the Dirac delta function,  readily reduces to 
$ \frac{1}{(2\pi)^3} \int  d {\bf k}  (2\pi)^3\delta_D({\bf k})  e^{- i {\bf k} \cdot {\bf r}} =   e^{- i {\bf 0} \cdot {\bf r}} = 1$. For a  system of a finite volume $V$, it is also customary to write
\begin{align}
\label{eq:kron}
 \int\limits_V  d{\bf r}   e^{ i {\bf k} \cdot {\bf r}} = V \delta_{{\bf k}, 0},
\end{align}
where $\delta_{i,j} = 1, i=j$ and $0, i\ne j$ is the Kronecker delta function.

\section{Direct and Reciprocal Lattices}

\label{sec:lat}
The position vector ${\bf r}$ of particles or atoms residing on a crystal lattice, the {\em direct lattice},  can be expressed as a linear combination, 
\begin{align*}
{\bf r} = x {\bf a} + y {\bf b} + z {\bf c},
\end{align*}
of the (direct) {\em  lattice vectors} ${\bf a}, {\bf b}, {\bf c}$, which are basis vectors of the unit cell with volume $V_{\rm cell} =  {\bf a} \cdot  ({\bf b} \times {\bf c}) $.   Generally,   ${\bf a}, {\bf b}, {\bf c}$ may not be orthogonal to each other and thus $x, y, z$ are not necessarily the projections of  {\bf r}  in a Cartesian coordinate system.
If particles coincide with lattice points, then $x, y, z$ are integers; if  partices are contained inside the unit cell, their coordinates $x, y, z$  can be fractions~\cite{sands1993}.

\begin{figure}
\centering
\includegraphics[width=0.4\textwidth]{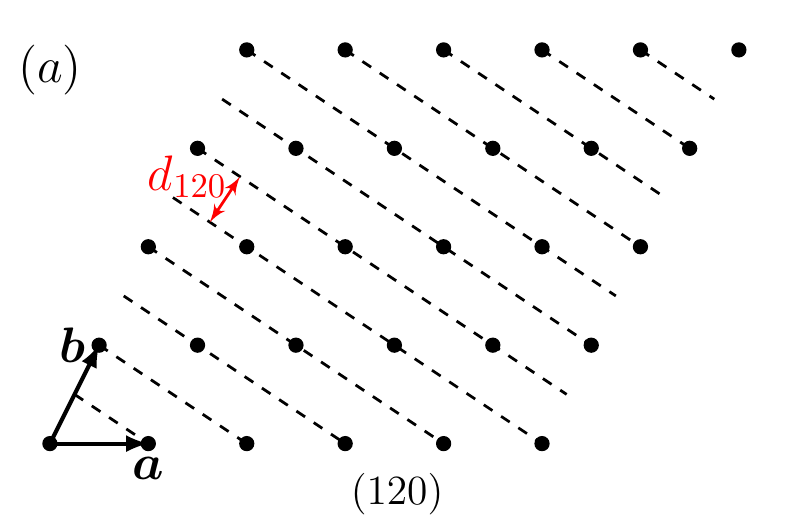}
\includegraphics[width=0.25\textwidth]{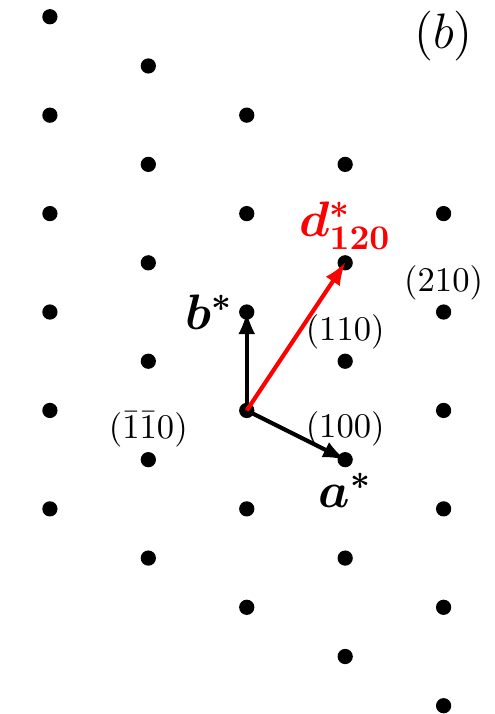}
\caption{(a) Direct and (b) reciprocal lattices. The $(120)$ planes (dashed lines) with interplanar distance $d_{120}$  in direct space correspond to the point denoted by the  vector ${\bf d}_{120}^*$  in the reciprocal space.  ${\bf d}_{120}^*$ is normal to the $(120)$  planes and  $| {\bf d}^*_{120} | = 1 / d_{120}$.  }
\label{fig:lattice}
\end{figure}

Particles on regular crystal lattices are situated on different families of parallel crystallographic planes, when viewed from different angles. Such parallel planes are denoted by  three integers $(hkl)$, the {\em Miller indices}, whose reciprocals are proportional to the intercepts of the planes with the three axes of the direct lattice. The spacing or distance, $d_{hkl}$, between  neighboring lattice planes in the family $(hkl)$ is a function of the Miller indices and lattice parameters (Fig.~\ref{fig:lattice}a). In the special case of an orthorhombic lattice,  
\begin{align*}
\frac{1}{d^2_{hkl}} = \frac{h^2}{a^2} + \frac{k^2}{b^2} +\frac{l^2}{c^2}.
\end{align*}

The {\em reciprocal lattice} is defined mathematically in a space spanned by the  {\em reciprocal lattice vectors} ${\bf a}^*, {\bf b}^*, {\bf c}^*$, which are related to the direct lattice vectors  by 
\begin{align*}
{\bf a}^* = ({\bf b} \times {\bf c}) /  V_{\rm cell}\\
{\bf b}^* = ({\bf c} \times {\bf a}) /  V_{\rm cell}\\
{\bf c}^* = ({\bf a} \times {\bf b}) /  V_{\rm cell}
\end{align*}
Since ${\bf a}^*$ is orthogonal to  $({\bf b}, {\bf c})$,   ${\bf b}^*$ is orthogonal to $({\bf a},  {\bf c})$  and $ {\bf c}^*$ is orthogonal to  $ ({\bf a}, {\bf b})$, 
\begin{align*}
{\bf a}^*   \cdot {\bf a} = 1, ~  {\bf a}^*   \cdot {\bf b} = 0, ~   {\bf a}^*   \cdot {\bf c} = 0,~ {\rm etc.}
\end{align*}
Note that,  in general,  ${\bf a}^*, {\bf b}^*, {\bf c}^*$ are not orthogonal to each other.
 Positions of reciprocal lattice points can be represented by vectors  of the form
\begin{align*}
 {\bf d}^*_{hkl} = h {\bf a}^* + k {\bf b}^* + l  {\bf c}^* 
 \end{align*} 
where $h,k,l$ are integers (Fig.~\ref{fig:lattice}b).

In crystallography, as the notation here implies, the physical meaning of the reciprocal lattice is related to  lattice planes in the direct space as follows~\cite{chen1986}:
\begin{itemize}
\item Each point with  a vector  $ {\bf d}^*_{hkl}$ on the reciprocal lattice represents a family of lattice planes with Miller indices $(hkl)$;
\item  The direction of  $ {\bf d}^*_{hkl}$ is perpendicular to (or normal to)  the lattice planes $(hkl)$;
\item The magnitude of $ {\bf d}^*_{hkl}$ is equal to the reciprocal of the interplanar spacing $d_{hkl}$, i.e $| {\bf d}^*_{hkl} | = 1 / d_{hkl}$.
\end{itemize}

 \end{appendix}

\bibliography{scattering.bib}

\end{document}